\documentclass[11pt]{article}
\input{psfig.sty}
\begin{document}

\title{
\textbf{Transfer matrix and Monte Carlo tests of critical exponents
in lattice models}
}

\author{J. Kaupu\v{z}s
\thanks{E--mail: \texttt{kaupuzs@latnet.lv}} \\
Institute of Mathematics and Computer Science, University of Latvia\\
29 Rainja Boulevard, LV--1459 Riga, Latvia}

\date{\today}

\maketitle

\begin{abstract}
The corrections to finite-size scaling in the critical two-point correlation
function $G(r)$ of 2D Ising model on a square
lattice have been studied numerically by means of exact transfer--matrix
algorithms. The systems of square geometry with periodic boundaries oriented
either along $\langle 10 \rangle$ or along $\langle 11 \rangle$ direction
have been considered, including up to 800 spins. The calculation of $G(r)$
at a distance $r$ equal to
the half of the system size $L$ shows the existence of an amplitude
correction $\propto L^{-2}$. A nontrivial correction $\propto L^{-0.25}$
of a very small magnitude also has been detected in agreement with
predictions of our recently developed GFD (grouping of Feynman diagrams)
theory. A refined analysis of the recent MC data for 3D Ising, $\varphi^4$,
and XY lattice models has been performed.
It includes an analysis of the partition function zeros of 3D Ising model,
an estimation of the correction--to--scaling exponent $\omega$ from the
Binder cumulant data near criticality,
as well as a study of the effective critical exponent $\eta$ and the
effective amplitudes in the asymptotic expansion of susceptibility at
the critical point. In all cases a refined analysis
is consistent with our (GFD) asymptotic values of the critical exponents
($\nu=2/3$, $\omega=1/2$, $\eta=1/8$ for 3D Ising model and 
$\omega=5/9$ for 3D XY model), while the actually accepted
"conventional" exponents are, in fact, effective exponents which are
valid for approximation of the finite--size scaling behavior of not too
large systems.
\end{abstract}


\vspace*{1ex}

{\bf Pacs:} 64.60.Cn, 68.18.Jk, 05.10.-a

\section{Introduction}

Since the exact solution of two--dimensional Ising model has
been found by Onsager~\cite{Onsager}, a study of various phase
transition models is of permanent interest. Nowadays, phase
transitions and critical phenomena is one of the most widely
investigated fields of physics~\cite{Sornette,BDT}.
Remarkable progress has been
reached in exact solution of two--dimensional models~\cite{Baxter}.
Recently, we have proposed~\cite{K1} a novel method based on
grouping of Feynman diagrams (GFD) in $\varphi^4$ model.
Our GFD theory allows to analyze the asymptotic solution
for the two--point correlation function at and near criticality,
not cutting the perturbation series. As a result the possible
values of exact critical exponents have been obtained~\cite{K1} for
the Ginzburg--Landau ($\varphi^4$) model with $O(n)$ symmetry,
where $n=1, 2, 3, \ldots$ is the dimensionality of the order
parameter. Our predictions completely agree with the
known exact and rigorous results in two dimensions~\cite{Baxter},
and are equally valid also in three dimensions. In~\cite{K1},
we have compared our results to some Monte Carlo (MC) simulations
and experiments~\cite{IS,SM,GA}. It has been shown~\cite{K1} that the
actually discussed MC data for 3D Ising~\cite{IS} and
$XY$~\cite{SM} models are fully consistent with our theoretical
predictions, but not with those of the perturbative renormalization
group (RG) theory~\cite{Wilson,Ma,Justin}. From the theoretical
and mathematical point of view, the invalidity of the conventional
RG expansions has been demonstrated in~\cite{K1}.
The current paper, dealing with numerical transfer-matrix analysis of
the two--point correlation function in 2D Ising model, as well as with the
analysis of MC data for the three--dimensional Ising, $\lambda \varphi^4$
and XY models, presents a more general confirmation that the correct values
of critical exponents are those predicted by the GFD theory.
Our estimations are based on the finite--size scaling theory,
which by itself is an attractive field of investigations~\cite{Chamati}
and has increasing importance in modern physics~\cite{BDT}.

\section{Critical exponents predicted by GFD theory} \label{sec:crex}

 Our theory provides possible values of exact critical exponents
$\gamma$ and $\nu$ for the $\varphi^4$ model whith $O(n)$
symmetry ($n$--component vector model) given by the Hamiltonian
\begin{equation} \label{eq:Ha}
H/T= \int \left[ r_0 \varphi^2({\bf x})
+ c (\nabla \varphi({\bf x}))^2
+ u \varphi^4({\bf x}) \right] d{\bf x} \; ,
\end{equation}
where $r_0$ is the only parameter depending on temperature $T$,
and the dependence is linear.  At the spatial
dimensionality $d=2, 3$ and $n=1, 2, 3, ...$ the critical
exponents are~\cite{K1}
\begin{eqnarray}
\gamma &=& \frac{d+2j+4m}{d(1+m+j)-2j} \label{eq:gamma} \; , \\
\nu &=& \frac{2(1+m)+j}{d(1+m+j)-2j} \label{eq:nu} \; ,
\end{eqnarray}
where $m \ge 1$ and $j \ge -m$ are integers. At $n=1$ we have
$m=3$ and $j=0$ to fit the known exact results for the
two--dimensional Ising model. As proposed in Ref.~\cite{K1},
in the case of $n=2$ we have $m=3$ and $j=1$, which yields in
three dimensions $\nu=9/13$ and $\gamma=17/13$.

In the present analysis the correction--to--scaling
exponent $\theta$ for the susceptibility is also relevant. The susceptibility
is related to the correlation function in the Fourier representation
$G({\bf k})$, i.~e., $\chi \propto G({\bf 0})$~\cite{Ma}. In the
thermodynamic limit, this relation makes sense at $T > T_c$, where
$T_c$ is the critical temperature.
According to our theory, $t^{\gamma} G({\bf 0})$ can be expanded in a Taylor
series of $t^{2 \nu -\gamma}$ at $t \to 0$.
In this case the reduced temperature $t$ is defined as
$t=r_0(T)-r_0(T_c) \propto T-T_c$.
Formally, $t^{2 \gamma - d \nu}$ appears as second expansion
parameter in the derivations in Ref.~\cite{K1}, but,
according to the final result represented by
Eqs.~(\ref{eq:gamma}) and~(\ref{eq:nu}),
$(2 \gamma - d \nu)/(2 \nu -\gamma)$ is a natural number.
Some of the expansion coefficients can be zero, so that in general we have
\begin{equation} \label{eq:Delta}
\theta=\ell \, (2 \nu -\gamma) \; ,
\end{equation}
where $\ell$ may have integer values 1, 2, 3, etc. One can expect
that $\ell=4$ holds at $n=1$ (which yields $\theta=1$ at $d=2$ and
$\theta=1/3$ at $d=3$) and the only nonvanishing
corrections are those of the order $t^{\theta}$, $t^{2 \theta}$,
$t^{3 \theta}$, since the known corrections to scaling for
physical quantities, such as magnetization or correlation length,
are analytical in the case of the two--dimensional Ising model.
Here we suppose that the confluent corrections become analytical,
i.~e. $\theta$ takes the value $1$, at $d=2$.
Besides, similar corrections to scaling are expected for
susceptibility $\chi$ and magnetization $M$ since both these
quantities are related to $G({\bf 0})$, i.~e.,
$\chi \propto G({\bf 0})$ and $M^2=\lim_{x \to \infty}
\langle \varphi({\bf 0}) \varphi({\bf x}) \rangle
= \lim_{V \to \infty} G({\bf 0})/V$
hold where $V=L^d$ is the volume and $L$ is the linear size of
the system. The above limit is meaningful at $L \to \infty$,
but $G({\bf 0})/V$ may be considered as a definition of $M^2$
for finite systems too. The latter means that corrections
to finite--size scaling for $\chi$ and $M$ are similar at $T=T_c$.
According to the scaling hypothesis and finite--size scaling
theory, the same is true for the discussed here corrections at $t \to 0$,
where in both cases ($\chi$ and $M$) the definition
$t= \mid r_0(T)-r_0(T_c) \mid$ is valid.
Thus, the expected expansion of the susceptibility $\chi$ looks
like $\chi = t^{-\gamma} \left( a_0+a_1 t^{\theta} +a_2 t^{2 \theta}
+ \cdots \right)$.

Our hypothesis is that $j=j(n)$ and $\ell=\ell(n)$
monotoneously increase with $n$ to fit the known exponents
for the spherical model at $n \to \infty$.
The analysis of the MC and experimental results here and in~\cite{K1}
enables us to propose that $j(n)=n-1$,
$\ell(n)=n+3$, and $m=3$ hold at least at $n=1,2$. These relations,
probably, are true also at $n \ge 3$.
This general hypothesis is consistent with the idea that
the critical exponents $\gamma$, $\nu$, and $\theta$
can be represented by some analytical functions of $n$ which are
valid for all natural positive $n$ and yield
$\eta=2-\gamma/\nu \propto 1/n$ rather than $\eta \propto 1/n^s$
with $s=2,3, \ldots$ ($s$ must be a natural number to avoid a
contradiction, i.~e., irrational values of $j(n)$ at natural $n$) at
$n \to \infty$. At these conditions, $j(n)$ and $\ell(n)$ are
linear functions of $n$ (with integer coefficients) such that
$\ell(n)/j(n) \to 1$ at $n \to \infty$, and $m$ is constant.
Besides, $j(1)=0$, $m(1)=3$, and $\ell(1)=4$ hold to coincide with the
known results at $n=1$.
Then, our specific choice is the best one among few possibilities
providing more or less reasonable agreement with the actually discussed
numerical an experimental results.

We allow that different $\ell$  values correspond to
the leading correction--to--scaling exponent for different
quantities related to $G({\bf k})$. The expansion of
$G({\bf k})$ by itself contains a nonvanishing term of order
$t^{2 \nu -\gamma} \equiv t^{\eta \nu}$ (in the form
$G({\bf k}) \simeq\linebreak t^{-\gamma} \left[ g({\bf k} t^{-\nu})
+ t^{\eta \nu} g_1({\bf k} t^{-\nu}) \right]$ whith
$g_1({\bf 0})=0$, since $\ell >1$ holds in the case of susceptibility)
to compensate the corresponding correction term (produced
by $c \left( \nabla \varphi \right)^2$) in the equation
for $1/G({\bf k})$ (cf.~\cite{K1}).
The latter means, e.~g., that the correlation
length $\xi$ estimated from an approximate ansatz like
$G({\bf k}) \propto 1/ \left[{\bf k}^2+ (1/\xi)^2 \right]$
used in~\cite{Janke,Ballesteros} also contains a correction
proportional to $t^{\eta \nu}$. Since $\eta \nu$ has a rather small value,
the presence of such a correction (and, presumably, also the higher order
corrections $t^{2 \eta \nu}$, $t^{3 \eta \nu}$, etc.) makes the above
ansatz unsuitable for an accurate numerical correction--to--scaling
analysis.

The correction $t^{\eta \nu}$ is related to the correction $L^{-\eta}$ 
in the finite--size scaling expressions at criticality.
 The existence of such a correction in the asymptotic
expansion of the critical real--space Green's (correlation) function
is confirmed by our results for the 2D~Ising model discussed
in Sec.~\ref{sec:result}.

  Our consideration can be generalized easily to the case
where the Hamiltonian parameter $r_0$ is a nonlinear analytical
function of $T$. Nothing is changed in the above expansions
if the reduced temperature $t$, as before, is defined by
$t=r_0(T)-r_0(T_c)$. However, analytical corrections to scaling appear
(and also corrections like\linebreak $(T-T_c)^{m+n \theta}$ with
integer $m$ and $n$) if $t$ is reexpanded in terms of $T-T_c$ at $T>T_c$.
The solution at the critical point remains unchanged, since the phase
transition occurs at the same (critical) value of $r_0$.

\section{Exact transfer matrix algorithms for calculation of the
 correlation function in 2D Ising model}
\label{sec:algorithm}

\subsection{Adoption of standard methods}

The transfer matrix method,
applied to analytical calculations on two--dimensional lattices, is well 
known~\cite{Onsager,Baxter}. However, no analytical methods exist for an exact 
calculation of the correlation function in 2D Ising model. This can be done
numerically by adopting the conventional transfer matrix
method and modifying it to reach the maximal result (calculation of as far 
as possible larger system) with minimal number of arithmetic operations,
as discussed further on.

\begin{figure}
\centerline{\psfig{figure=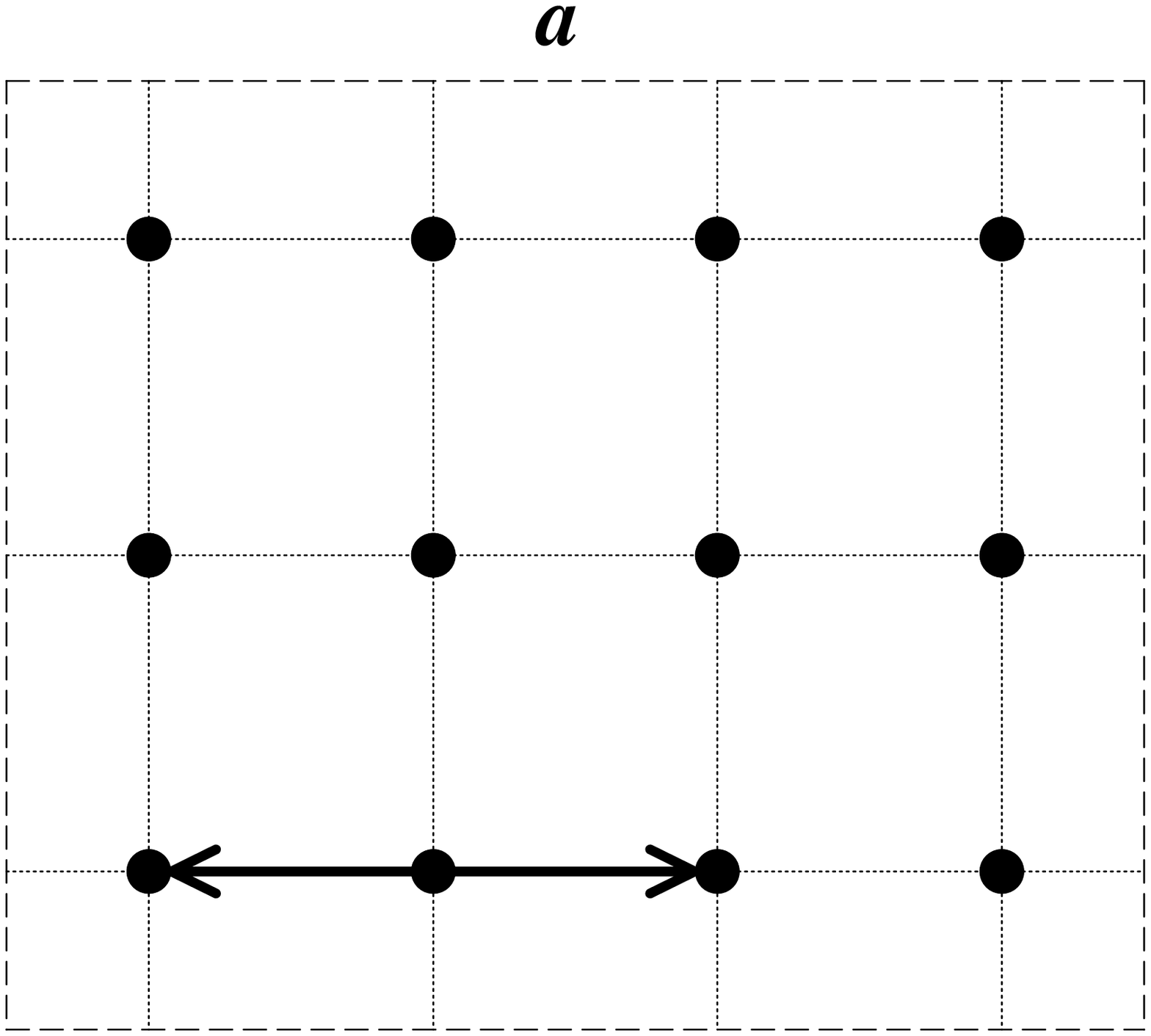,width=6.5cm,height=5.5cm,angle=-90}
	\hspace*{5ex}
            \psfig{figure=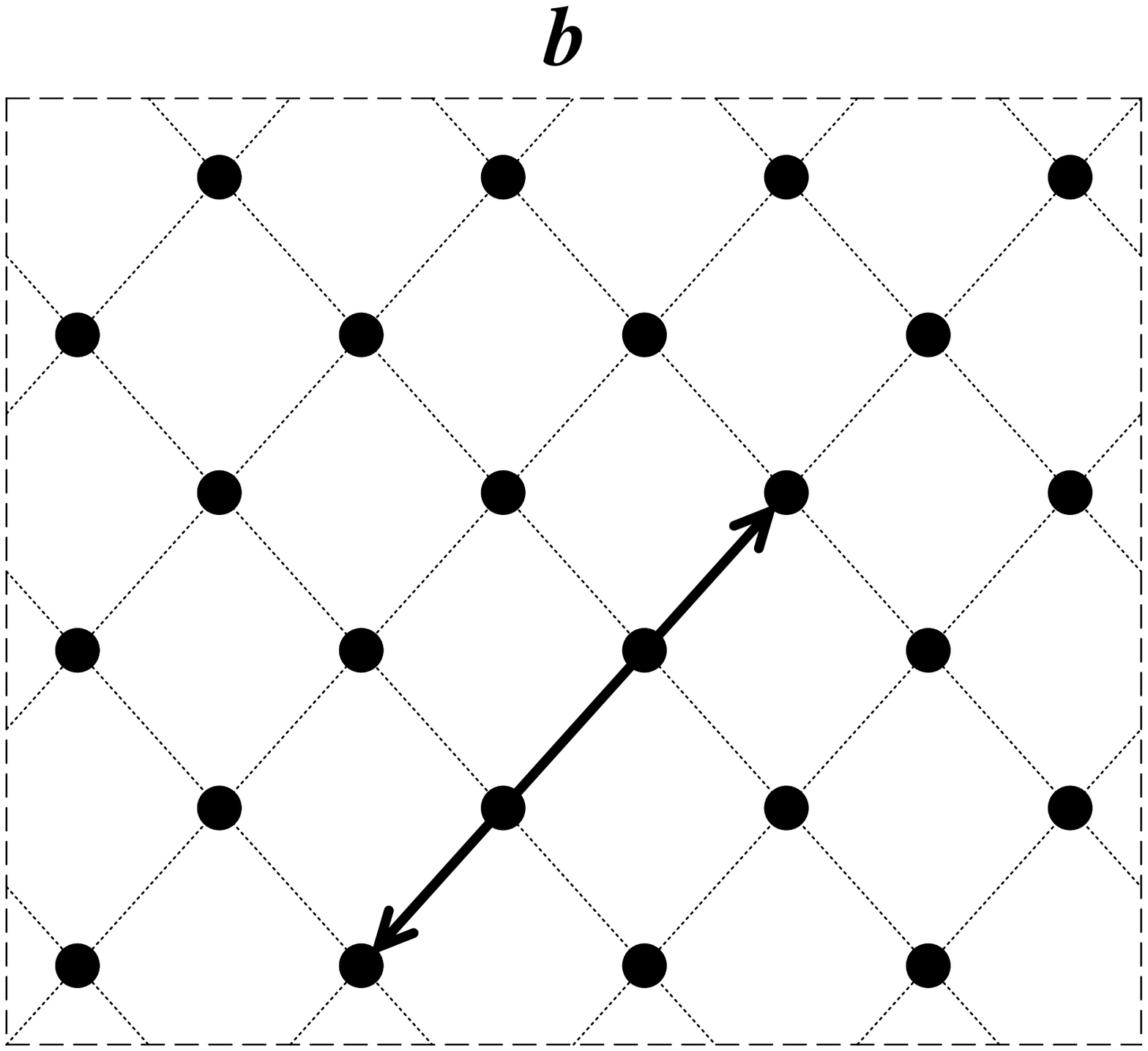,width=6.5cm,height=5.5cm,angle=-90}}
\caption{\small Illustrative examples of the lattices with
dimensions $N \times L$~(a) and $\sqrt{2} N \times \sqrt{2} L$~(b)
with periodic boundary conditions along the dashed lines. 
The correlation function has been calculated in the $\langle 10 \rangle$
crystallographic direction, as indicated by the arrows.}
\label{lattice}
\end{figure}

 We consider the two--dimensional Ising model where spins are located
either on the lattice of dimensions $N \times L$, illustrated in 
Fig.~\ref{lattice}a, or on the lattice of dimensions $\sqrt{2} N \times
\sqrt{2} L$, shown in Fig.~\ref{lattice}b. 
The periodic boundaries
are indicated by dashed lines. In case (a) we have $L$ rows,
and in case (b) -- $2L$ rows, each containing $N$ spins. 
Fig.~\ref{lattice} shows an illustrative example with $N=4$ and $L=3$. 
Let us $\sigma(k)$ be the spin variable ($\pm 1$) in the $k$--th
node of the first row. Here nodes are numbered sequently from 
left to right, and rows -- from bottom to top.
Our method can be used to calculate the 
correlation (Greens) function between any two spins on the lattice.
As an example we consider the Greens function $G(r)$ in 
$\langle 1 0 \rangle$ crystallographyc direction, 
indicated in Fig.~\ref{lattice} by arrows, i.~e.,
\begin{eqnarray}
G(r) &=& \langle \sigma(k) \sigma(k+r) \rangle 
\hspace*{9.5ex}: \hspace{5ex} \mbox{case (a)} \label{eq:cora} \\
G(r) &=& \langle \sigma(k) \sigma'(k+\Delta(r)) \rangle
\hspace*{5ex}: \hspace{5ex} \mbox{case (b)} \label{eq:corb} \;.
\end{eqnarray}  
Here $\sigma'$ refers to the $(1+r)$--th row. It has a shift in the
argument $k$ by $\Delta(r)=r/2$ for even and $\Delta(r)=(r-1)/2$ for odd $r$.
It is supposed that
$\sigma(k+N) \equiv \sigma(k)$ holds according to the periodic
boundary conditions.

For convenience, first we consider an application of the transfer matrix
method to calculation of the partition function
\begin{equation}
Z= \sum\limits_{ \{\sigma_k \} } 
\exp \left( \beta \sum\limits_{\langle i,j \rangle} 
\sigma_i \sigma_j \right) \;,
\end{equation}
where the summation runs over all the possible spin
configurations $\{ \sigma_k \}$, and the argument of the exponent
represents the Hamiltonian of the system including summation over
all the neighbouring spin pairs $\langle i,j \rangle$ of the given 
configuration $\{ \sigma_k \}$; 
parameter $\beta$ is the coupling constant.
Let us consider lattice~(a) in Fig.~\ref{lattice}, but containing $n$ rows
without periodic boundaries along the vertical axis and without interaction
between spins in the upper row. We define the
$2^N$--component vector ${\bf r}_n$ such that the $i$--th
component of this vector represents the contribution to the partition 
function provided by the $i$--th spin configuration of the upper
row. Then we have an obvious recurrence relation
\begin{equation}
{\bf r}_{n+1}=T \, {\bf r}_n \;,
\label{eq:recur_a}
\end{equation}
where $T$ is the transfer matrix with the elements
\begin{equation}
T_{ij}= \exp \left( \beta \sum\limits_k \left[ \sigma(k) \right]_j
\left[ \sigma(k+1) \right]_j + 
\beta \sum\limits_k \left[ \sigma(k) \right]_j 
\left[ \sigma(k) \right]_i \right) \;.
\label{eq:trm}
\end{equation}
Here $\left[ \sigma(k) \right]_i$ is the spin variable in the
$k$--th position in a row provided that the whole set of 
spin variables of this row forms the $i$--th configuration.
The first and the second sum in~(\ref{eq:trm}) represent the
Boltzmann weights for the spin interaction in the $n$--th row,
and between the $n$--th and $(n+1)$--th rows, respectively.
In this case, ${\bf r}_1$ is the vector with components 
$\left( {\bf r}_1 \right)_j =1$. If we set
\begin{equation} 
\left( {\bf r}_1 \right)_j = \delta_{j,i} \;,
\label{eq:delta}
\end{equation}
then the components
of the resulting vector ${\bf r}_n={\bf r}_n^{(i)}$ give us the 
partial contributions
to the partition function corresponding to a fixed, i.~e. $i$--th, 
configuration of the first row. The periodic boundary
conditions along the vertical axis means that the $(L+1)$--th
row must be identical to the first one, i.~e., we have to take
the $i$--th component of the vector ${\bf r}_{L+1}^{(i)}$ and make
the summation over $i$ to get the partition function $Z$ of the 
originally defined lattice in Fig.~\ref{lattice}a.
Note that the missing Boltzmann weights 
for the interaction between spins in the $(L+1)$--th row
are already included in the first row.
By virtue of~(\ref{eq:recur_a}) and~(\ref{eq:delta}), we arrive
to the well known expression~\cite{Baxter,Huang}
\begin{equation}
Z= \sum\limits_i \left( {\bf r}_{L+1}^{(i)} \right)_i = 
\mbox{Trace} \left(T^L \right)= \sum\limits_i \lambda_i^L \;,
\end{equation}
where $\lambda_i$ are the eigenvalues of the transfer matrix $T$.
An analogous expression for the lattice in Fig.~\ref{lattice}b reads
\begin{equation}
Z= \sum\limits_i \left( {\bf r}_{2L+1}^{(i)} \right)_i = \mbox{Trace} 
\left( \left[ T_2 T_1 \right]^L \right) 
\;,
\end{equation}
where the vectors ${\bf r}_n$ obey the reccurence relation
\begin{equation}
{\bf r}_{n+1}=T_{1,2} \, {\bf r}_n 
\label{eq:recur_b}
\end{equation}
similar to~(\ref{eq:recur_a}), but with different transfer
matrices $T_1$ and $T_2$ for odd and even row numbers $n$, respectively.
They include the Boltzmann weights for the interaction between 
two subsequent (odd--even or even--odd, respectively) rows, i.~e.,
\begin{equation}
\left( T_{1,2} \right)_{ij} = \exp \left( \beta \sum\limits_k
\left[ \sigma(k) \right]_i \left\{ \left[ \sigma(k) \right]_j
+ \left[ \sigma(k \pm 1) \right]_j \right\} \right) \;.
\end{equation}

The actual scheme can be easily adopted to calculate the
correlation functions~(\ref{eq:cora}) and~(\ref{eq:corb}).
Namely, $G(x)$ is given by the statistical average $Z' / Z$,
where the sum $Z'$ is calculated in the same way as $Z$, but
including the corresponding product of spin variables, which
implies the following replacements:
\begin{eqnarray}
\left( {\bf r}_1 \right)_j = \delta_{j,i} \Rightarrow
\left( {\bf r}_1 \right)_j = \delta_{j,i} \,
\left( N^{-1} \sum\limits_{\ell=1}^N \left[ \sigma(\ell) \right]_i
\left[ \sigma(\ell+x) \right]_i \right)&& : \mbox{case~(a)} 
\label{eq:Zpa} \\
\left( {\bf r}_{x+1}^{(i)} \right)_j \Rightarrow 
\left( {\bf r}_{x+1}^{(i)} \right)_j \times
\left( N^{-1} \sum\limits_{\ell=1}^N \left[ \sigma(\ell) \right]_i
\left[ \sigma(\ell+ \Delta(x)) \right]_j \right)&& : \mbox{case~(b)} \;.
\label{eq:Zpb}
\end{eqnarray} 
The index $i$, entering in the sums 
$Z'= \sum_i \left( {\bf r}_{L+1}^{(i)} \right)_i$ [case~(a)] and
$Z'= \sum_i \left( {\bf r}_{2L+1}^{(i)} \right)_i$ [case~(b)],
refers to the current configuration of the first row. 
These equations are obtained by an averaging in (\ref{eq:cora}) 
and~(\ref{eq:corb}) over all the equivalent $k$ values. 
Such a symmetrical form allows to reduce the amount of numerical calculations:
due to the symmetry we need the summation over only $\approx 2^N/N$
nonequivalent configurations of the first row instead of the total
number of $2^N$ configurations.

\subsection{Improved algorithms}

The number of the required arithmetic operations can be further reduced 
if the recurrence relations~(\ref{eq:recur_a}) and~(\ref{eq:recur_b})
are split into $N$ steps of adding single spin. To formulate this
in a suitable way, let us first number all the $2^N$ spin configurations
$\{ \sigma(1); \sigma(2); \cdots ; \sigma(N-1); \sigma(N) \}$ by an
index $i$ as follows:
\begin{equation}
\begin{array}{@{i \: = \:}l@{\hspace{2ex} : \hspace{3ex} \{}*{6}{c}@{\}}}
1 & -1; & -1; & \cdots; & -1; & -1; & -1 \\
2 & -1; & -1; & \cdots; & -1; & -1; & +1 \\
3 & -1; & -1; & \cdots; & -1; & +1; & -1 \\
4 & -1; & -1; & \cdots; & -1; & +1; & +1 \\
\multicolumn{7}{c}\dotfill \\
2^N & +1; & +1; & \cdots; & +1; & +1; & +1 
\end{array}
\label{eq:numbering}
\end{equation}
We remind that the sequence $\left[ \sigma(k) \right]_i$ with
$k=1, 2, \ldots, N$ corresponds to the numbers in the $i$--th row.
The spin variables in~(\ref{eq:numbering}) change just like
the digits of subsequent integer numbers in the binary counting system.

\begin{figure}
\centerline{\psfig{figure=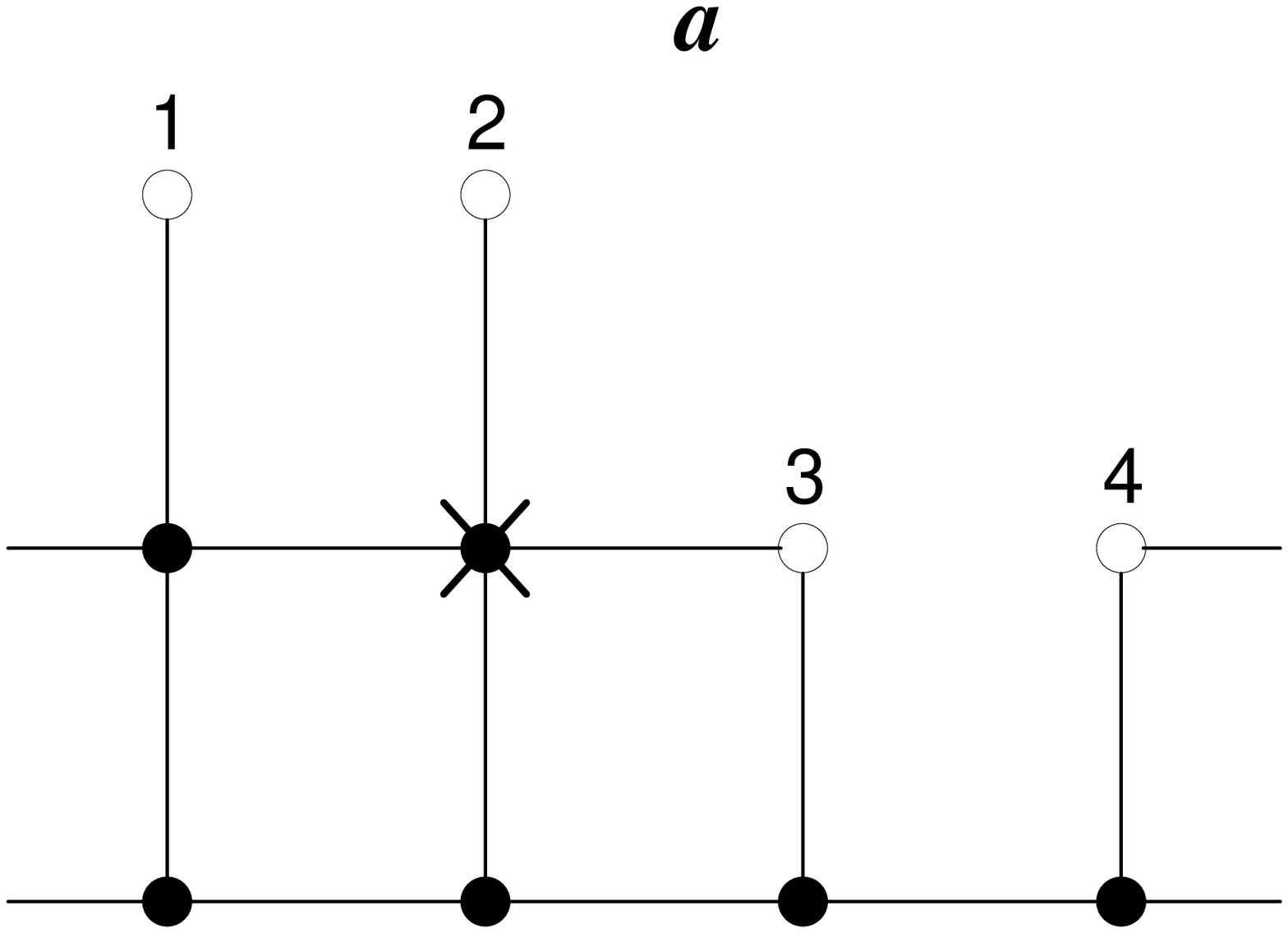,width=6.5cm,height=5.5cm,angle=-90}
	\hspace*{5ex}
            \psfig{figure=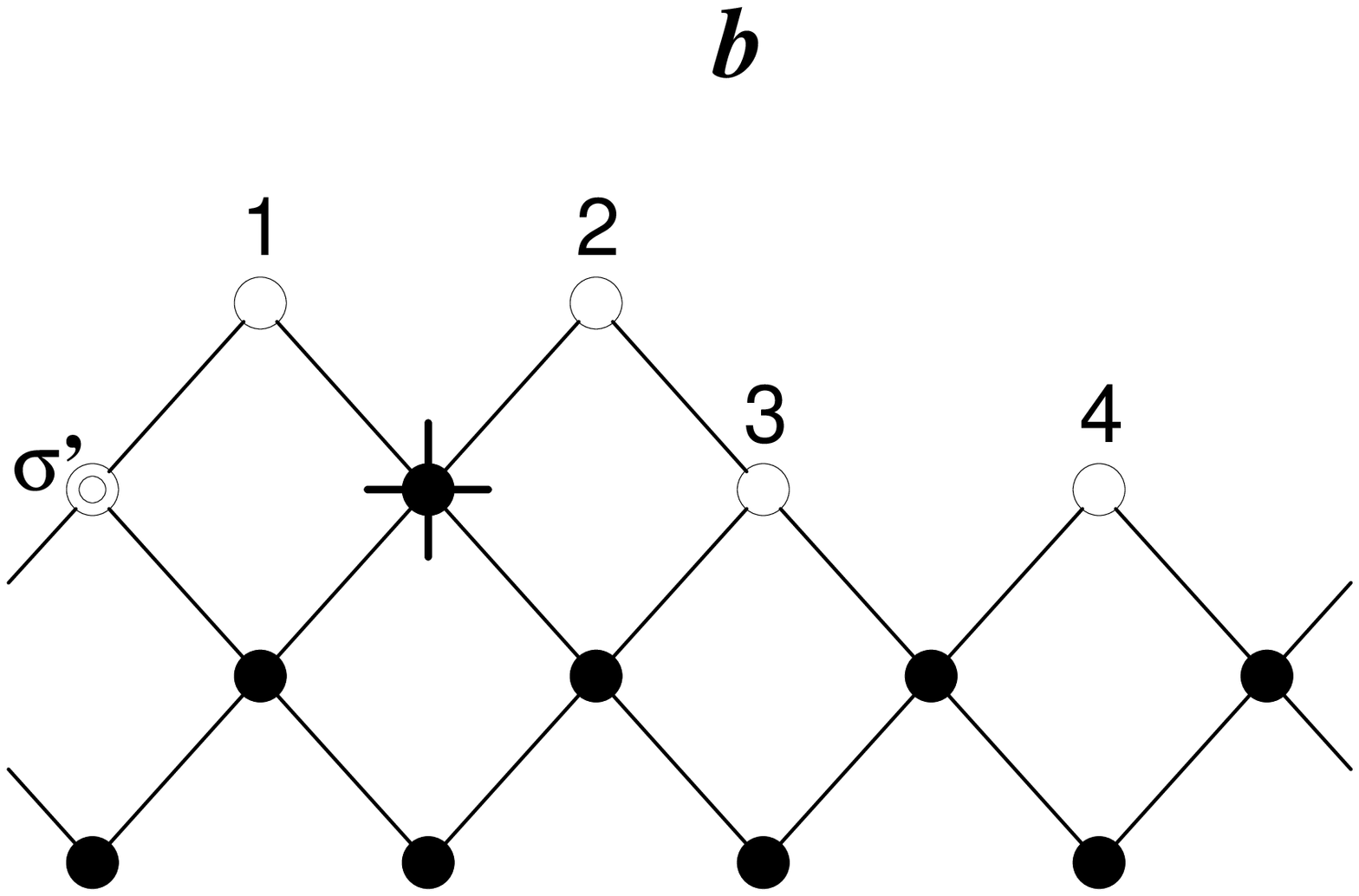,width=6.5cm,height=5.5cm,angle=-90}}
\vspace*{-2ex}
\caption{\small Schematic pictures illustrating the algorithms of 
calculation for the lattices $a$ and $b$ introduced in Fig.~\ref{lattice}.}
\label{latt}
\end{figure}

 Consider now a lattice where $n$ rows are completed, while the
$(n+1)$--th row contains only $\ell$ spins where $\ell < N$, as 
illustrated in Fig.~\ref{latt} in both cases~(a) and~(b) taking
as an example $N=4$. We consider the partial contribution 
$\left( {\bf r}_{n+1,\ell} \right)_i$ (i.~e., $i$--th component of vector
${\bf r}_{n+1,\ell}$) 
in the partition function $Z$ (or $Z'$) provided by a fixed ($i$--th) 
configuration of the set of $N$ upper spins. 
These are the sequently numbered spins
shown in Fig.~\ref{latt} by empty circles.
For simplicity, we have droped 
the index denoting the configuration of the first row. 
In case~(b), the spin depicted
by a double--circle has a fixed value $\sigma'$.
In general, this spin is the nearest bottom--left neighbour of the first 
spin in the upper row. According to this, one has to distinguish 
between odd and even $n$: $\sigma'$ refers either to the first 
(for odd $n$), or to the $N$--th (for even $n$) spin of the $n$--th row. 
It is supposed that the Boltzmann weights are included corresponding
to the solid lines in Fig.~\ref{latt} connecting the spins.
In case~(a) the weights responsible
for the interaction between the upper numbered spins are not included.
Obviously, for a given $\ell>1$, ${\bf r}_{n+1,\ell}$
can be calculeted from ${\bf r}_{n+1,\ell-1}$ via summation over
one spin variable, marked in Fig.~\ref{latt} by a cross.
In case~(a) it is true also for $\ell=1$, whereas in case~(b)
this variable has fixed value $\sigma'$ at $\ell=1$. In the latter case 
the summation over $\sigma'$ is performed at the last step when
the $(n+1)$--th row is already completed. These manipulations enable
us to represent the recurrence relation~(\ref{eq:recur_a}) as 
\begin{equation}
{\bf r}_{n+1}=T \, {\bf r}_n \equiv
\widetilde W_N \, \widetilde W_{N-1} \cdots \widetilde W_2 \,
\widetilde W_1 \, {\bf r}_n 
\label{eq:rec_a}
\end{equation}
with
\begin{equation}
\widetilde W_{\ell} = \sum\limits_{\sigma= \pm 1} W_{\ell} (\sigma) \;,
\label{eq:wa}
\end{equation}
where the componets of the matrices $W_{\ell}(\sigma)$ are given by
\begin{eqnarray}
\left( W_1(\sigma) \right)_{ij} &=& \delta \left( j,j_1(\sigma ,1,i) \right)
\cdot \exp \left( \beta \, \sigma \left\{ \left[ \sigma(1) \right]_i  
+ \left[ \sigma(2) \right]_i + \left[ \sigma(N) \right]_i \right\} \right) 
\nonumber \\
\left( W_{\ell}(\sigma) \right)_{ij} &=& \delta \left( j,j_1(\sigma,\ell,i) 
\right)\cdot \exp \left( \beta \, \sigma \left\{ \left[ \sigma(\ell) \right]_i
+ \left[ \sigma(\ell+1) \right]_i \right\} \right) \hspace{1ex}: 
\hspace{1ex} 1< \ell <N \nonumber \\
\left( W_N(\sigma) \right)_{ij} &=& \delta \left( j,j_1(\sigma,N,i) \right)
\cdot \exp \left( \beta \, \sigma \left[ \sigma(N) \right]_i \right) \;.
\label{eq:a}
\end{eqnarray}
Here $\delta(j,k)$ is the Kronecker symbol and
\begin{equation}
j_1(\sigma,\ell,i)= i+ \left( \sigma- \left[ \sigma(\ell) \right]_i \right)
\, 2^{N-\ell-1}
\label{eq:j1}
\end{equation}
are the indexes of the old configurations containing $\ell-1$ spins in 
the $(n+1)$--th row depending on the value $\sigma$ of the spin
marked in Fig.~\ref{latt}a by a cross, as well as on the index $i$ of the 
new configuration with $\ell$ spins in the $(n+1)$--th row, as consistent
with the numbering~(\ref{eq:numbering}).

The above equations~(\ref{eq:rec_a}) to~(\ref{eq:a}) refers to case~(a). 
In case~(b) we have
\begin{equation}
{\bf r}_{n+1}=T_{1,2} \, {\bf r}_n \equiv \sum\limits_{\sigma'= \pm 1}
\widetilde W_N^{(1,2)} \, \widetilde W_{N-1}^{(1,2)} \cdots 
\widetilde W_2^{(1,2)} \, W_1^{(1,2)}(\sigma') \, {\bf r}_n \;,
\label{eq:rec_b}
\end{equation}
where $\widetilde W_{\ell}^{(1,2)}$ are the matrices 
\begin{equation}
\widetilde W_{\ell}^{(1,2)} = 
\sum\limits_{\sigma= \pm 1} W_{\ell}^{(1,2)} (\sigma) \;.
\label{eq:wb}
\end{equation}
Here indexes 1 and 2 refer to odd and even row numbers $n$, respectively,
and the components of the matrices $W_{\ell}^{(1,2)}(\sigma)$ are 
\begin{equation}
\left( W_{\ell}^{(1,2)}(\sigma) \right)_{ij} 
= \delta \left( j,j_{1,2}(\sigma,\ell,i)
\right) \cdot \exp \left( \beta \left[ \sigma(\ell) \right]_i 
\left\{ \sigma+ \left[ \sigma(\ell+1) \right]_i \right\} \right) \;,
\end{equation}
where $\left[ \sigma(N+1) \right]_i \equiv \sigma'$ and the index 
$j_1(\sigma,\ell,i)$ is given by~(\ref{eq:j1}). For the other index
we have
\begin{eqnarray}
j_2(\sigma,1,i) &=& 2i - 2^{N-1} \left( \left[ \sigma(1) \right]_i +1 \right) 
+ \frac{1}{2} (\sigma-1) \nonumber \\ 
j_2(\sigma,\ell,i) &=& j_1(\sigma,\ell,i) \hspace{2ex}: 
\hspace{2ex} \ell \ge 2 \;.
\end{eqnarray}

Note that the matrices $\widetilde W_{\ell}$ and 
$\widetilde W_{\ell}^{(1,2)}$ have only two nonzero elements in
each row, so that the number of the arithmetic operations required
for the construction of one row of spins via subsequent calculation
of the vectors ${\bf r}_{n+1,\ell}$ increases like 
$2N \cdot 2^N$ instead of $2^{2N}$ operations necassary for a straightforward 
calculation of the vector $T {\bf r}_n$. Taking into account
the above discussed symmetry of the first row, the computation 
time is proportional to $2^{2L}L$ for both $L \times L$ (a)
and $\sqrt{2} L \times \sqrt{2} L$ (b) lattices in Fig.~\ref{lattice}
with periodic boundary conditions.
 
\subsection{Application to different boundary conditions}
\label{subsec:anti}

The developed algorithms can be easily extended
to the lattices with antiperiodic boundary conditions. The latter implies
that $\sigma(N+1)=-\sigma(N)$ holds for each row, and similar condition
is true for each column. We can consider also the mixed boundary
conditions: periodic along the horizontal axis and antiperiodic along
the vertical one, or vice versa. To replace the periodic boundary conditions
with the antiperiodic ones we need only to change the sign of the
corresponding products of the spin variables on the boundaries.
Consider, e.~g., the case~(a) in Fig.~\ref{lattice}. The change of the
boundary conditions along the vertical axis means that the first term
in the argument of the exponent in each of the Eqs.~(\ref{eq:a}) changes 
the sign for the last row, i.~e., when $n=L$. 
The same along the horizontal axis implies
that the term $\left[ \sigma(N) \right]_i$ in the equation for
$\left( W_1(\sigma) \right)_{ij}$ changes the sign. In this case, however,
the symmetry with respect to the configurations of the first row is partly 
broken and, therefore, we need summation over a larger number of  
nonequivalent configurations.

\section{Transfer matrix study of critical Greens function 
and corrections to scaling in 2D Ising model}
\label{sec:correc}

\subsection{General scaling arguments}
\label{subsec:scaling}

It is well known that in the thermodynamic limit the real--space Greens 
function of the Ising model behaves like $G(r) \propto r^{2-d-\eta}$ at 
large distances $r \to \infty$ at the critical point $\beta=\beta_c$, 
where $\eta$ is the critical exponent having the value $\eta=1/4$ in 
two dimensions ($d=2$). Based on our transfer matrix algorithms developed
in Sec~\ref{sec:algorithm}, here we test the finite--size scaling and, 
particularly, the corrections to scaling at criticality.

In~\cite{K1} the critical correlation function in 
the Fourier representation, i.~e. $G({\bf k})$ at $T=T_c$, 
has been considered for the $\varphi^4$ model. In this case the minimal 
value of the wave vector magnitude $k$ is related to the linear system size 
$L$ via $k_{min}=2 \pi/ L$. In analogy to the consideration in
Sec.~5.2 of~\cite{K1}, one expects that $k/k_{min}$ is an essential
finite--size scaling argument, corresponding to $r/L$ in the real space.
In the Ising model at $r \sim L$ one has to take into account
also the anisotropy effects, so that the expected finite--size scaling
relation for the real--space Greens function at the critical point
$\beta=\beta_c$ reads
\begin{equation}
G(r) \simeq r^{2-d-\eta} \, f(r/L) \hspace{5ex}:
\hspace{3ex} r \to \infty \;, L \to \infty \;,
\label{eq:scal}
\end{equation}
where the scaling function $f(z)$ depends also on the
crystallographic orientation of the line connecting the correlating spins, 
as well as on the orientation of the periodic boundaries. 
A natural extension of~(\ref{eq:scal}), including the corrections to scaling, 
is
\begin{equation}
G(r) = \sum\limits_{\ell \ge 0} r^{-\lambda_{\ell}} \, f_{\ell}(r/L) \;,
\label{eq:scale}
\end{equation}
where the term with $\lambda_0 \equiv d-2+\eta$ is the leading one,
whereas those with the subsequently increasing exponents $\lambda_1$, 
$\lambda_2$, etc., represent the corrections to scaling.
By a substitution $f_{\ell}(z)=z^{\lambda_{\ell}} f'(z)$, 
the asymptotic expansion~(\ref{eq:scale}) transforms to
\begin{equation}
G(r) = f'_0(r/L) \,  L^{-\lambda_0} \left( 1 + \sum\limits_{\ell \ge 1} 
L^{-\omega_{\ell}} \, \tilde f_{\ell}(r/L) \right) \;,
\label{eq:scalel1}
\end{equation}
where $\tilde f_{\ell}(z) = f_{\ell}'(z)/f_0(z)$ and
$\omega_{\ell}=\lambda_{\ell}-\lambda_0$ are the correction--to--scaling
exponents.

\begin{figure}
\centerline{\psfig{figure=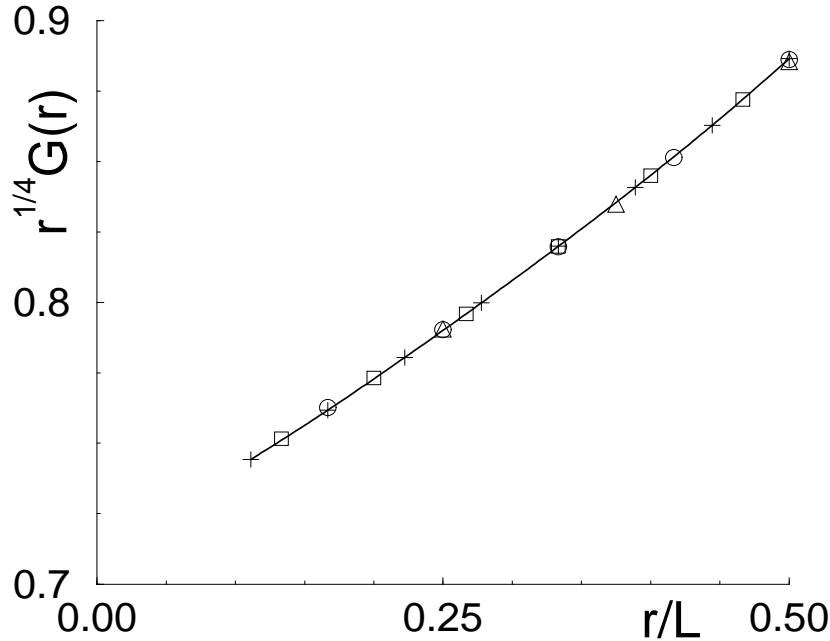,width=11cm,height=8.5cm,angle=-90}}
\caption{\small Test of the finite--size scaling relation 
$G(r) \simeq r^{-1/4} f(r/L)$ in 2D Ising model on
a square $L \times L$ lattice with periodic boundary conditions.
The values of the scaling function $f(r/L)$ estimated at
$L=8$ (triangles), $L=12$ (circles), $L=15$ (squares), and $L=18$ (pluses)
well lie on a common smooth line showing that corrections to this
scaling relation are small.
}
\label{fig:scaling}
\end{figure}

We have tested the scaling relation~(\ref{eq:scal}) in 2D Ising model
by using the exact transfer matrix algorithms in Sec.~\ref{sec:algorithm}.
Consider, e.~g., the correlation function in $\langle 10 \rangle$
crystallographic direction in the specific case~(a) discussed in 
Sec.~\ref{sec:algorithm}. According to~(\ref{eq:scal}), all points
of $f(r/L)=r^{1/4}G(r)$ corresponding to large enough values
of $r$ and $L$ should well fit a common smooth line, as it is actually
observed in Fig~\ref{fig:scaling} at $2 \le r \le L/2$ and
$L=8$, $12$, $15$, and $18$. This example shows that the corrections 
to~(\ref{eq:scal}) are rather small, since the deviations from the
spline curve constructed at $L=18$ are very small at relatively
small $L$ values and even at $r=2$.
In general, our calculations provide a strong numerical evidence
of correctness of the asymptotic expansion~(\ref{eq:scalel1}): 
at a given ratio $r/L$, the correlation function is described by
an expansion in $L$ powers with a really striking accuracy.
Namely, our numerical analysis, which is sensitive to a variation
of $G(r)$ in the fourteens digit, does not reveal
an inconsistency with this asymptotic expansion.  

\subsection{Correction--to--scaling analysis for the
$L \times L$ lattice}
\label{subsec:ll}

Based on the scaling analysis in Sec.~\ref{subsec:scaling},
here we discuss the corrections to scaling for the lattice
in Fig.~\ref{lattice}a. We have calculated the correlation 
function $G(r)$ at a fixed ratio $r/L=0.5$ in $\langle 10 \rangle$ direction, 
as well as at $r/L= 0.5 \sqrt{2}$ in $\langle 11 \rangle$ direction at 
$L=2, 4, 6, \ldots$ with an aim to identify the correction exponents 
in~(\ref{eq:scalel1}). Note that in the latter case the 
replacement~(\ref{eq:Zpb}) is valid for $G(\sqrt{2}x)$ 
(where $x=1, 2, 3, \ldots$) with the only difference that $\Delta(x)=x$. 

Let us define the effective correction--to--scaling exponent 
$\omega_{eff}(L)$ in 2D Ising model via the solution of the equations
\begin{equation}
\tilde L^{1/4}G(r=const \cdot \tilde L) = a + b \, \tilde L^{-\omega_{eff}}
\label{eq:omfit}
\end{equation}
at $\tilde L=L, \, L+\Delta L, \, L+ 2 \Delta L$ with respect to three 
unknown quantities $\omega_{eff}$, $a$, and $b$. 
According to~(\ref{eq:scalel1}), where $\lambda_0=\eta=1/4$,
such a definition gives us the leading correction--to--scaling
exponent $\omega$ at $L \to \infty$, i.~e., $\lim_{L \to \infty} 
\omega_{eff}(L)=\omega$. 

\begin{table}
\caption{\small The correlation function $G(r=c \cdot L)$ in $\langle 10 \rangle$ 
($c=0.5$) and $\langle 11 \rangle$ ($c=0.5 \sqrt{2}$) crystallographic 
directions  vs the linear size $L$ of the lattice~(a) in Fig.~\ref{lattice}, 
and the corresponding effective exponents $\omega_{eff}(L)$ and $\widetilde \omega(L)$.}
\label{tab1}
\vspace*{2ex}
\begin{center}
\begin{tabular}{|c|c|c|c|c|c|}
\hline
& \multicolumn{2}{|c|}{\rule[-2.5mm]{0mm}{7mm} direction $\langle 10 \rangle$} 
& \multicolumn{3}{|c|}{direction $\langle 11 \rangle$} \\ \cline{2-6}
\raisebox{1.5ex}{L} 
  & \rule[-3mm]{0mm}{7.5mm}
    $G(0.5 L)$       & $\omega_{eff}(L)$ & $G(0.5 \sqrt{2}L)$ & $\omega_{eff}(L)$ & $\widetilde \omega(L)$ \\ \hline
2 & 0.84852813742386 & 2.7366493         & 0.8                & 1.8672201         &                        \\ 
4 & 0.74052044609665 & 2.9569864         & 0.71375464684015   & 2.2148707         &                        \\
6 & 0.67202206468538 & 1.8998036         & 0.65238484475089   & 2.1252078         &                        \\
8 & 0.62605120856389 & 1.5758895         & 0.60935351016910   & 2.0611362         & 1.909677               \\
10& 0.59238112628953 & 1.6617494         & 0.57724041054810   & 2.0351831         & 1.996735               \\
12& 0.56615525751968 & 1.7774398         & 0.55200680271678   & 2.0232909         & 2.002356               \\
14& 0.54485584658226 & 1.8542943         & 0.53141907668442   & 2.0167606         & 2.001630               \\
16& 0.52703456475995 &                   & 0.51414720882560   &                   &                        \\
18& 0.51178753041103 &                   & 0.49934511003360   &                   &                        \\ \hline
\end{tabular}
\end{center}
\end{table}

The calculated values of $G(r= c \cdot L)$ in the $\langle 10 \rangle$ and 
$\langle 11 \rangle$ crystallographic directions [in case~(a)] with $c=0.5$ 
and $c=0.5 \sqrt{2}$, respectively, and the corresponding
effective exponents $\omega_{eff}(L)$, determined at $\Delta L=2$, are given 
in Tab.~\ref{tab1}. In both cases the effective exponent $\omega_{eff}(L)$
seems to converge to a value about $2$. Besides, in the second case the
behavior is more smooth, so that we can try someway to extrapolate
the obtained sequence of $\omega_{eff}$ values (column 5 in Tab.~\ref{tab1})
to $L= \infty$. For this purpose we have considered the ratio of two
subsequent increments in $\omega_{eff}$,
\begin{equation}
r(L)= \frac{\omega_{eff}(L+\Delta L)- \omega_{eff}(L)}
           {\omega_{eff}(L)- \omega_{eff}(L-\Delta L)} \;.
\label{eq:rl}
\end{equation}
A simple analysis shows that $r(L)$ behaves like
\begin{equation}
r(L) = 1 - \Delta L \cdot (\omega'+1) L^{-1} +o \left( L^{-2} \right)
\label{eq:rapp}
\end{equation}
at $L \to \infty$ if $\omega_{eff}(L) = \omega + o \left( L^{-\omega'}
\right)$ holds with an exponent $\omega'>1$. 
The numerical data in Tab.~\ref{tab1} show that Eq.~(\ref{eq:rapp})
represents a good approximation for the largest values of $L$ at $\omega'=2$.
It suggests us that the leading and the subleading correction exponents 
in~(\ref{eq:scalel1}) could be $\omega \equiv \omega_1=2$ and $\omega_2=4$, 
respectively. Note that $\omega_{eff}(L)$ can be defined with a shift in
the argument. Our specific choice ensures
the best approximation by~(\ref{eq:rapp}) at the actual finite $L$ values.

Let us now assume that the values of $\omega_{eff}(L)$ are known up to
$L=L_{max}$. Then we can calculate from~(\ref{eq:rl}) the $r(L)$ values
up to $L=L_{max}-\Delta L$ and make a suitable ansatz like
\begin{equation}
r(L)= 1 - 3 \Delta L \cdot L^{-1} + b \, L^{-2} 
\hspace{3ex} \mbox{at} \hspace{2ex} L \ge L_{max}
\label{eq:rapp1}
\end{equation}
for a formal extrapolation of $\omega_{eff}(L)$ to $L=\infty$.
This is consistent with~(\ref{eq:rapp}) where $\omega'=2$. The
coefficient $b$ is found by matching the result to the precisely
calculated value at  $L=L_{max}-\Delta L$. 
The subsequent values of $\omega_{eff}(L)$, calculated
from~(\ref{eq:rl}) and~(\ref{eq:rapp1}) at $L>L_{max}$, converge
to some value $\widetilde \omega(L_{max})$ at $L \to \infty$.
If the leading correction--to--scaling exponent $\omega$
is $2$, indeed, then the extrapolation result
$\widetilde \omega(L_{max})$ will tend to $2$ at $L_{max} \to \infty$
irrespective to the precise value of $\omega'$.

As we see from Tab.~\ref{tab1}, the values of $\widetilde \omega(L)$ 
come remarkably closer to $2$ as compared to $\omega_{eff}(L)$, which seems
to indicate that $\omega=2$. However, as we have discussed in
Sec.~\ref{sec:crex}, there should be a nontrivial correction 
in~(\ref{eq:scalel1}) with $\omega= \eta=1/4$. The fact that this
nontrivial correction does not manifest itself in the above numerical
analysis can be understood assuming that this correction term
has a very small amplitude. 

\subsection{Correction--to--scaling analysis for the
$\sqrt{2} L \times \sqrt{2} L$ lattice}
\label{sec:result}

To test the existence of nontrivial corrections to scaling, 
as proposed at the end of Sec.~\ref{subsec:ll}, here we make the 
analysis of the correlation function $G(r)$ in $\langle 10 \rangle$
direction on the $\sqrt{2} L \times \sqrt{2} L$ lattice shown
in Fig.~\ref{lattice}b. 
The advantage of case~(b) in Fig.~\ref{lattice}
as compared to case~(a) is that $\sqrt{2}$ times larger lattice corresponds
to the same number of the spins in one row.
Besides, in this case
we can use not only even, but all lattice sizes to evaluate the
exponent $\omega$ from calculations of $G(r=L)$, which means
that it is reasonable to use the step $\Delta L=1$ to evaluate
$\omega_{eff}$ and $\widetilde \omega(L)$ from Eqs.~(\ref{eq:omfit}),
(\ref{eq:rl}) and~(\ref{eq:rapp1}). The results, are given in Tab.~\ref{tab2}.
\begin{table}
\caption{\small The correlation function $G(r=L)$ in $\langle 10 \rangle$ 
crystallographic direction and the effective exponents 
$\omega_{eff}(L)$ and $\widetilde \omega(L)$ vs the linear size $L$ of 
the lattice~(b) in Fig.~\ref{lattice}.}
\label{tab2}
\vspace*{2ex}
\begin{center}
\begin{tabular}{|c|c|c|c|}
\hline
\rule[-3mm]{0mm}{7.5mm}
L & $G(L)$             & $\omega_{eff}(L)$ & $\widetilde \omega(L)$ \\ \hline
2 & 0.8                &                   &                        \\
3 & 0.7203484812087670 &                   &                        \\
4 & 0.6690636562097066 &                   &                        \\
5 & 0.6321925914229602 &                   &                        \\
6 & 0.6037455936471098 &                   &                        \\
7 & 0.5807668304926868 &                   &                        \\
8 & 0.5616046762441826 & 2.066235298       &                        \\
9 & 0.5452468033693456 & 2.043461090       &                        \\
10& 0.5310294874153481 & 2.030235674       & 1.996772124            \\
11& 0.5184950262041604 & 2.022130104       & 1.999333324            \\  
12& 0.5073151480587211 & 2.016864947       & 1.999941357            \\
13& 0.4972468711401118 & 2.013265826       & 2.000036957            \\
14& 0.4881056192765374 & 2.010701166       & 2.000040498            \\
15& 0.4797481011874659 & 2.008811505       & 2.000044005            \\
16& 0.4720609977942179 & 2.007380630       & 2.000053415            \\
17& 0.4649532511721054 & 2.006272191       & 2.000063984            \\
18& 0.4583506666254706 & 2.005396785       & 2.000073711            \\
19& 0.4521920457268738 &                   &                        \\ 
20& 0.4464263594840965 &                   &                        \\ \hline
\end{tabular}
\end{center}
\end{table}

It is evident from Tab.~\ref{tab2} that the extrapolated values of the
effective correction exponent, i.~e. $\widetilde \omega(L)$, come surprisingly
close to $2$ at certain $L$ values. Besides, the ratio of increments 
$r$~[cf.~Eq.~(\ref{eq:rl})] in this case is well approximated by~(\ref{eq:rapp1}),
as consistent with existence of a correction term in~(\ref{eq:scalel1})
with exponent $4$. On the other hand, we can see from Tab.~\ref{tab2} that 
$\Delta \widetilde \omega(L)= \widetilde \omega(L) - 2$ tends to increase in 
magnitude at $L>13$. 
\begin{figure}
\centerline{\psfig{figure=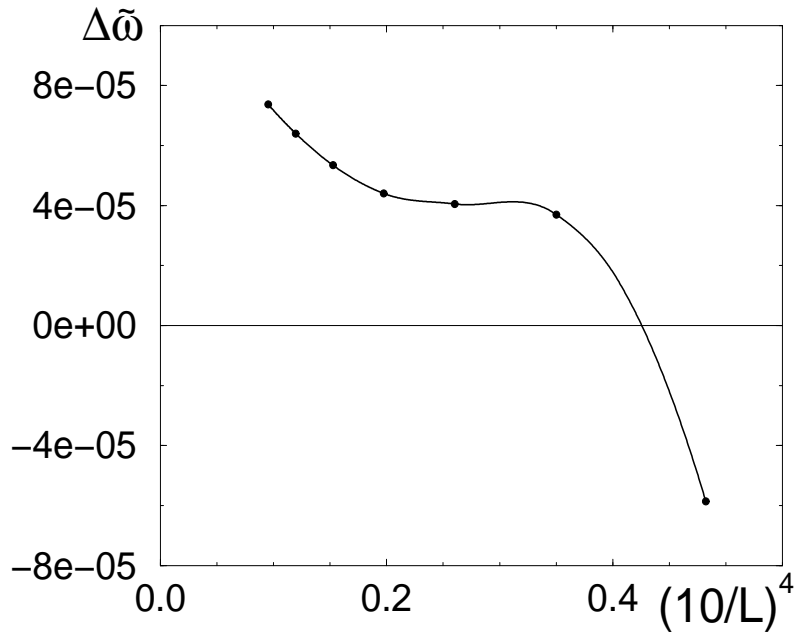,width=11cm,height=8.5cm,angle=-90}}
\caption{\small The deviation of the extrapolated effective exponent
$\Delta \widetilde \omega(L)=\widetilde \omega(L)-2$ as a 
function of $L^{-4}$. The extrapolation has been made by using the
calculated $G(r)$ values in Tab.~\ref{tab2} up to the size $L+2$.
A linear convergence to zero would be expected in absence of
any correction term with exponent $\omega<2$.}
\label{fig:dev}
\end{figure}
We have illustrated this systematic and smooth
deviation in Fig.~\ref{fig:dev}. The only reasonable explanation of this behavior
is that the expansion~(\ref{eq:scalel1}) necessarily contains the exponent
$2$ and, likely, also the exponent $4$, and at the same time it contains
also a correction of a very small amplitude with $\omega<2$. The latter
explains the increase of $\Delta \widetilde \omega(L)$. Namely, the
correction to scaling for $L^{1/4}G(L)$ behaves like
$const \cdot L^{-2} \left[ 1 + o \left( L^{-2} \right) 
+ \varepsilon \, L^{2-\omega} \right]$ with $\varepsilon \ll 1$,
which implies a slow crossover of the effective exponent $\omega_{eff}(L)$
from the values about $2$ to the asymptotic value $\omega$. 
Besides, in the region where $\varepsilon \, L^{2-\omega} \ll 1$ holds,
the effective exponent behaves like
\begin{equation}
\omega_{eff}(L) \simeq 2 + b_1 L^{2-\omega} + b_2 L^{-2} \;,
\label{eq:oeff}
\end{equation} 
where $b_1 \ll 1$ and $b_2$ are constants.
By using the extrapolation of $\omega_{eff}$ with $\omega'=2$ in~(\ref{eq:rapp})
and~(\ref{eq:rapp1}), we have compensated the effect of the correction term 
$b_2 L^{-2}$. Besides, by matching the amplitude $b$ in~(\ref{eq:rapp1}) we
have compensated also the next trivial correction term $\sim L^{-3}$ in
the expansion of $\omega_{eff}(L)$. It means that the extrapolated exponent 
$\widetilde \omega(L)$ does not contain these expansion terms, i.~e., we have
\begin{equation}
\widetilde \omega(L) = 2 + b_1 L^{2-\omega} + \delta \widetilde \omega(L) \;,
\label{eq:omext}
\end{equation} 
where $\delta \widetilde \omega(L)$ represents a remainder term. It 
includes the trivial corrections like $L^{-4}$, $L^{-5}$, etc.,
and also subleading nontrivial corrections,
as well as corrections of order $\left(\varepsilon \, L^{2-\omega} \right)^2$,
$\left(\varepsilon \, L^{2-\omega} \right)^3$, etc.,
neglected in~(\ref{eq:oeff}). According to the latter,
Eq.~(\ref{eq:omext}) is meaningless in the thermodynamic limit 
$L \to \infty$, but it can be used
to evaluate the correction--to--scaling exponent $\omega$ from the
transient behavior at large, but not too large values of $L$ where
$b_1 L^{2-\omega} \ll 1$ holds. In our example the latter condition is well
satisfied, indeed. 

\begin{figure}
\centerline{\psfig{figure=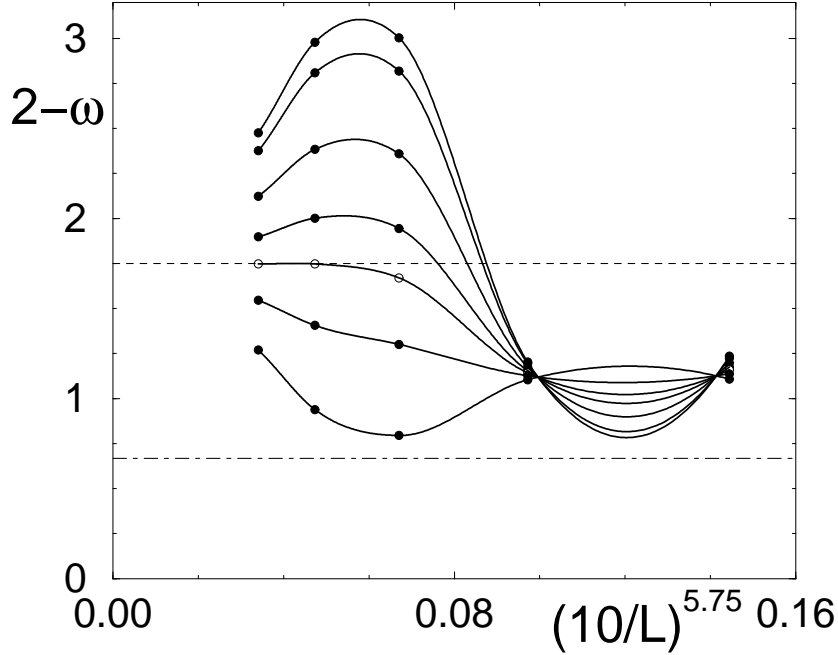,width=11cm,height=8.5cm,angle=-90}}
\caption{\small The exponent $2-\omega$ estimated from~(\ref{eq:lincom})
at different system sizes.
From top to bottom (if looking on the left hand side):
$\alpha = 0, 1, 3.5, 5.75, 7.243, 9.25, 12$. The results at
the optimal $\alpha$ value $7.243$ are shown by empty circles. The dashed line
indicates our theoretical asymptotic value $2-\omega =1.75$, whereas the
dot--dashed line -- that proposed in~\cite{BF}.}
\label{fig:om}
\end{figure}

Based on~(\ref{eq:omext}), we have estimated
the nontrivial correction--to--scaling exponent $\omega$ by using the
data of $\widetilde \omega(L)$ in Tab~\ref{tab2}. We have used two different ansatzs
\begin{equation}
2-\omega_1 (L)= \ln \left[ \Delta \widetilde \omega(L) / 
\Delta \widetilde \omega(L-1) \right] / \ln [ L/(L-1)] 
\end{equation}
and
\begin{equation}
2-\omega_2 (L)= L \, \left[ \Delta \widetilde \omega (L) - \Delta \widetilde 
\omega (L-1) \right] / \Delta \widetilde \omega (L) \;,
\end{equation}
as well as the linear combination of them
\begin{equation}
\omega(L)=(1-\alpha) \; \omega_1(L) + \alpha \; \omega_2(L)
\label{eq:lincom}
\end{equation}
containing a free parameter $\alpha$. We have $\omega(L)=\omega_1(L)$
at $\alpha=0$ and $\omega(L)=\omega_2(L)$ at $\alpha=1$. In general,
the effective exponent $\omega(L)$ converges to the same result $\omega$
at arbitrary value of $\alpha$, but at some values the convergence is better.
The results for $2-\omega(L)$ vs $L^{\omega-6}$ at different $\alpha$ values
are represented in Fig.~\ref{fig:om} by a set of curves. 
In this scale the convergence
to the asymptotic value would be linear 
(within the actual region where $L \gg 1$ and
$b_1 \, L^{2-\omega} \ll 1$ hold) for $\alpha=0$ at the condition
$\delta \widetilde \omega(L) \propto L^{-4}$.
We have choosen the scale of $L^{-5.75}$, as it is consistent with our
theoretical prediction in Sec.~\ref{sec:crex} that $\omega=1/4$. Nothing is 
changed essentially if we use slightly different scale as, e.~g., 
$L^{-14/3}$ consistent with the correction--to--scaling exponent $\omega=4/3$ 
proposed in~\cite{BF}.
As we see from Fig.~\ref{fig:om}, all curves tend to merge at our asymptotic
value $2 -\omega=1.75$ shown by a dashed line. The optimal
value of $\alpha$ is defined by the condition that the last two estimates 
$\omega(17)$ and $\omega(18)$ agree with each other. It occurs
at $\alpha=7.243$, and the last two points lie just on our theoretical line.  

It is interesting to compare our results with those of the
high temperature (HT) series analysis in~\cite{Addendum}. 
The authors of~\cite{Addendum}
have found ``almost by inspection'' a correction with exponent
$\omega=9/4$ in the asymptotic expansion of the susceptibility $\chi$.
If such a correction exists in the susceptibility, it must be present
also in the correlation function due to the relation 
$\chi= \sum_{\bf x} G({\bf x}_1-{\bf x})$. Surprisingly, our 
extremely accurate calculations by exact algorithms have not
revealed such a correction. Our analysis shows that nontrivial
corrections exist, indeed, in the correlation function of 2D Ising
model, but they are so extremely small and strongly masked by trivial 
corrections like $L^{-2}$ and $L^{-4}$ that they could not be
detected by approximate methods like HT series expansion.

\subsection{Comparison to the known exact results and estimation
of numerical errors}
\label{sec:comex}

We have carefully checked our algorithms comparing the results
with those obtained via a straightforward counting of all spin
configurations for small lattices, as well as comparing the 
obtained values of the partition function to those calculated
from the known exact analytical expressions. Namely, an
exact expression for the partition function of a finite--size
2D lattice on a torus with arbitrary coupling constants between 
each pair of neighbouring spins has been reported in~\cite{Bednorz} 
obtained by the loop counting method and represented
by determinants of certain transfer matrices. In the standard 2D 
Ising model with only one common coupling constant $\beta$ these 
matrices can be diagonalized easily, using the standard 
techniques~\cite{Landau}. Besides, the loop counting method
can be trivially extended to the cases with antiperiodic
or mixed boundary conditions discussed in 
Sec.~\ref{subsec:anti}. It is necessary only to mention that
each loop gets an additional factor $-1$ when it winds round
the torus with antiperiodic boundary conditions. 
We consider the partition functions
$Z_{pp} \equiv Z$, $Z_{aa}$, $Z_{ap}$, $Z_{pa}$.
In this notation the first index refers to the horizontal
or $x$ axis, and the second one -- to the vertical or
$y$ axis of a lattice illustrated in Fig.~\ref{lattice}a;
$p$ means periodic and $a$ -- antiperiodic boundary conditions.
As explained above, the standard methods leads to the
following exact expressions:
\begin{eqnarray}
Z_{pp} &=& \left( Q_1+Q_2+Q_3-Q_0 \right)/ \, 2 \nonumber \\
Z_{ap} &=& \left( Q_0+Q_1+Q_3-Q_2 \right)/ \, 2 \nonumber \\
Z_{pa} &=& \left( Q_0+Q_1+Q_2-Q_3 \right)/ \, 2 \label{eq:zz} \\
Z_{aa} &=& \left( Q_0+Q_2+Q_3-Q_1 \right)/ \, 2 \nonumber 
\end{eqnarray}
where $Q_0$ is the partition function represented by the sum of 
the closed loops on the lattice, 
as consistent with the loop counting method in~\cite{Landau}, 
whereas  $Q_1$, $Q_2$, and $Q_3$
are modified sums with additional factors
$\exp(\Delta x \cdot i \pi/N + \Delta y \cdot i \pi/L)$, 
$\exp(\Delta x \cdot i \pi/N)$, and 
$\exp(\Delta y \cdot i \pi/L)$, respectively, 
related to each change of coordinate $x$ by $\Delta x = \pm 1$,
or coordinate $y$ by $\Delta y = \pm 1$ when making a loop. 
The standard manipulations~\cite{Landau} yield
\begin{eqnarray}
&&Q_i = 2^{NL} \prod\limits_{q_x, \, q_y} \left[ \cosh^2 (2 \beta)
-\sinh(2 \beta)  \right. \\
&&\left. \times \left( \cos \left[ q_x+ 
\left( \delta_{i,1}+\delta_{i,2} \right) \frac{\pi}{N} \right]
+ \cos \left[ q_y+ \left( \delta_{i,1}+\delta_{i,3} \right) 
\frac{\pi}{L} \right] \right) \right]^{1/2} \nonumber \;,
\label{eq:qi}
\end{eqnarray}
where the wave vectors $q_x=(2 \pi/N) \cdot n$ and 
$q_y=(2 \pi/L) \cdot \ell $ run over all the values
corresponding to $n=0, 1, 2, \ldots , N-1$
and  $\ell=0, 1, 2, \ldots , L-1$.
In the case of the periodic boundary conditions,
each loop of $Q_0$ has the sign $(-1)^{m+ab+a+b}$~\cite{Bednorz},
where $m$ is the number of intersections, $a$ is
the number of windings around the torus in $x$ direction, and 
$b$ -- in $y$ direction. The correct result for $Z_{pp}$ 
is obtained if each of the loops has the sign $(-1)^m$.
In all other cases, similar relations 
are found easily, taking into account the above defined
additional factors. Eqs.~(\ref{eq:zz}) are then obtained
by finding such a linear combination of quantities $Q_i$ which ensures
the correct weight for each kind of loops.

All our tests provided a perfect agreement between the
obtained values of the Greens functions $G(r)$ 
(a comparison between straightforward calculations and
our algorithms), as well as between partition
functions for different boundary conditions
(a comparison between our algorithms and Eq.~(\ref{eq:zz})). 
The relative discrepancies were extremely small (e.~g., $10^{-15}$),
obviously, due to the purely numerical inaccuracy.

We have used the double--precision FORTRAN programs. 
The main source of the inaccuracy in our 
calculations is the accumulation of numerical errors
during the summation of long sequences of numbers, i.~e.,
during the sumation over all the nonequivalent configurations
of the first row of spins. To eliminate the error for
the largest lattice $L=20$, we have split the summation in several
parts in such a way that a relatively small part, 
including only the first 10~000 configurations from the total number
of 52~487 nonequivalent ones, gives the main contribution
to $Z$ and $Z'$.
The same trick with splitting in two approximately
equal parts 
has been used at $L=19$, as well.
By comparing the summation results with different
splitings, we have concluded that a systematical error in
$G(r)$ at $L=20$ could reach the value about $3 \cdot 10^{-15}$. 
The calculations at $L=18$ and $19$ have been performed with
approximately the same accuracy. 
The systematical errors in subsequent $G(r)$ values tend to
compensate in the final result for $\omega(L)$. The resulting
numerical errors in Fig.~\ref{fig:om} are about $0.02$ or $0.03$,
i.~e., approximately within the symbol size.

\section{Analysis of the partition function zeros
in 3D Ising model}  \label{sec:zeros}

In this section we discuss the recent MC results~\cite{ADH}
for the complex zeros of the partition function of the
three--dimensional Ising model. Namely, if the coupling $\beta$
is a complex number, then the partition sum has zeros at
certain complex values of $\beta$ or $u= e^{-\beta}$.
The nearest to the real positive axis values $\beta_1^0$ and
$u_1^0$ are of special interest. Neglecting the second--order
corrections, $u_1^0$ behaves like
\begin{equation}
u_1^0 = u_c + A\, L^{-1/\nu} +B\, L^{-(1/\nu)-\omega}
\end{equation}
at large $L$, where $u_c=e^{-\beta_c}$ is the critical value
of $u$, $A$ and $B$ are complex constants, and $\omega$ is the
correction--to--scaling exponent. According to the known
solution given in~\cite{Landau,Brout}, the
partition function zeros correspond to complex values of
$\, \sinh (2 \beta) \,$ located on a unit circle
in the case of 2D Ising model, so that $A$ is purely imaginar.
This solution, however, is only asymptotically exact
at $L \to \infty$. Nevertheless, based on an analysis of the exact
expression Eq.~(\ref{eq:zz}) we conclude that the statement 
$Re \, A=0$ is
correct. This fact is obvious in the case of Brascamp--Kunz boundary
conditions~\cite{JK,JK1}. The latter
means that the critical behavior of real and imaginary
parts of $u_1^0 - u_c$ essentially differ from each other,
i.~e., $Re \left( u_1^0 -u_c \right) \propto L^{-(1/\nu) -\omega}$
and $Im \left( u_1^0 \right) \propto L^{-1/\nu}$
(where, in this case of $d=2$, $\nu=\omega=1$) at $L \to \infty$.
The MC data of~\cite{ADH}, in fact, provide a good evidence that
the same is true in three dimensions.
 
 Based on MC data for the partition function zeros in 3D Ising model,
the authors of Ref.~\cite{ADH} have searched the way how to confirm the
already known estimates for $\nu$. Their treatment, however, is
rather doubtful. First, let us mention that, in contradiction to
the definition in the paper, $u_1^0$ values
listed in Tab.~I of~\cite{ADH} are not equal to $e^{-\beta_1^0}$
(they look like $e^{-4\beta_1^0}$).
Second, the fit to a theoretical ansatz for $\mid u_1^0(L) - u_c \mid$,
Eq.~(6) in~\cite{ADH}, is unsatisfactory. This ansatz contains a
mysterious parameter $a_3$. If we compare Eqs.~(5) and (6)
in~\cite{ADH}, then we see immediately that
$a_3 \equiv (1/\nu)+\omega$. At the same time, the obtained estimate
for $a_3$, i.~e. $a_3 =4.861(84)$, is completely inconsistent with
the values of $(1/\nu)+\omega$, about $2.34$, which follow from
authors own considerations. Our prediction, consistent with
the correction--to--scaling analysis in Sec.~\ref{sec:crex}
(and with $\ell=4$ in~(\ref{sec:crex}) to coincide with the known
exact result at $d=2$), is $\nu=2/3$ and $\omega=1/2$, i.~e.,
$(1/\nu)+\omega=2$.

\begin{figure}
\centerline{\psfig{figure=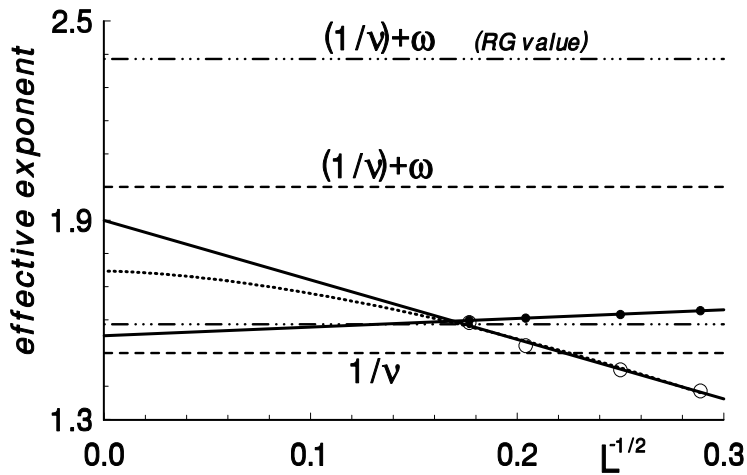,width=11cm,height=8.5cm}}
\vspace{-9.5ex}
\caption{\small Effective critical exponents for the real (empty circles)
and the imaginary (solid circles) part of complex
partition--function--zeros of 3D Ising model depending on
$L^{-1/2}$, where $L$ is the linear size of the system.
Solid lines show the linear least--squares fits. 
The asymptotic values from our theory are indicated by horizontal
dashed lines, whereas those of the RG theory -- by
dot--dot--dashed lines. A selfconsistent extrapolation within
the RG theory corresponds to the tiny dashed line.}
\label{zeros}
\end{figure}

  To obtain a more complete picture, we have considered separately
the real part and the imaginary part of $u_1^0-u_c$. We have calculated
$u_1^0$ from $\beta_1^0$ data listed in Tab.~I of~\cite{ADH} and
have estimated the effective critical exponents $y_{eff}'(L)$
and $y_{eff}''(L)$, separately
for $Re \left( u_1^0-u_c \right)$ and $Im \left( u_1^0 \right)$,
by fitting these quantities to an ansatz $const \cdot L^{-y_{eff}'}$
and $const \cdot L^{-y_{eff}''}$, respectively,
at sizes $L$ and $L/2$. The value of $u_c$ consitent with
the estimation of the critical coupling in~\cite{HV},
$\beta_c \simeq 0.2216545$, has been used.
The results are shown in Fig.~\ref{zeros}.
As we see, $y_{eff}'$ (empty circles) claims to increase
above $y_{eff}''$ (solid circles) when $L$ increases.
This is a good numerical evidence that, like in the two--dimensional
case, the asymptotic values are\linebreak
$y'=\lim_{L \to \infty} y_{eff}'(L)=(1/\nu)+\omega$ and
$y''=\lim_{L \to \infty} y_{eff}''(L)=1/\nu$.
 According to our theory, the actual plots in the $L^{-1/2}$ scale
are linear at $L \to \infty$, as consistent with the expansion
in terms of $L^{-\omega}$. The linear least--squares fits are shown
by solid lines. The zero intercepts $1.552$ and $1.899$ are in approximate
agreement with our theoretical values $1.5$ and $2$ indicated by horizontal
dashed lines. The relatively small discrepancy, presumably, is due
to the extrapolation errors and inaccuracy in the simulated data.
The result for $y'$ is affected by the error in $\beta_c$ value.
However, this effect is negligibly small.
Assuming a less accurate value $\beta_c=0.221659$, consistent with
the estimations in~\cite{SA,FL}, we obtain $y'=1.914$.

The behavior of $y_{eff}'$ is rather inconsistent with
the RG predictions. On the one hand, $y_{eff}'$ claims to increase
above $y_{eff}''$ and also well above the RG value
of $1/\nu$ (the lower dod--dot--dashed line at $1.586$), and, on the other
hand, the extrapolation yields $y'$ value ($1.899$) which is remarkably
smaller than
$(1/\nu)+\omega \simeq 2.385$ (the upper dot--dot--dashed line)
predicted by the RG theory. For selfconsistency,
we should use the linear extrapolation in the scale of $L^{-\omega}$ with
$\omega=0.799$ (the RG value). However, this extrapolation
(tiny dashed line in Fig.~\ref{zeros}), yielding
$y' \simeq 1.747$, does not solve the problem
in favour of the RG theory.

\begin{figure}
\centerline{\psfig{figure=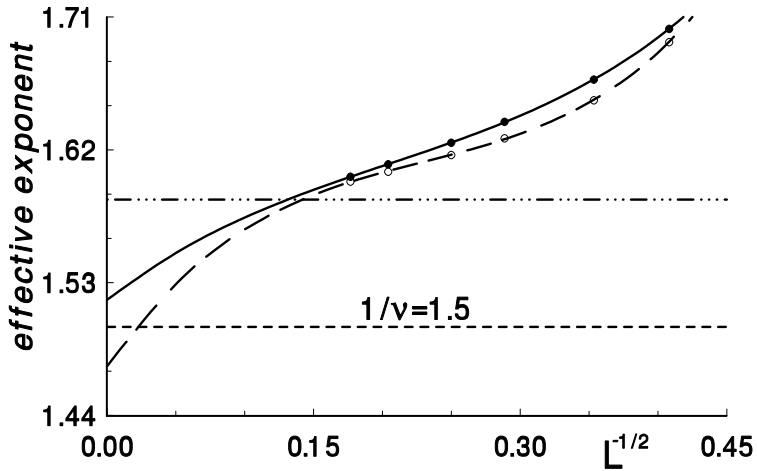,width=11cm,height=8.5cm}}
\vspace{-9.5ex}
\caption{\small Effective critical exponent $y_{eff}''(L)$
for the imaginary part of the complex partition--function--zeros
as a function of $L^{-1/2}$, where $L$ is the linear size of the system.
The empty circles correspond to $\phi = \exp(-\beta)$, whereas the solid
circles to $\phi = \sinh( 2 \beta)$. The corresponding least--squares
fits $y_{eff}''(L)= 1.4734 + 1.3321 L^{-1/2} - 4.7587 L^{-1}
+ 6.8894 L^{-3/2}$ and
$y_{eff}''(L)= 1.5184 + 0.7271 L^{-1/2} - 2.0309 L^{-1}
+ 3.3095 L^{-3/2}$ are shown by long--dashed line
and solid line, respectively. 
Our asymptotic value $y''=1/\nu=1.5$ is indicated by horizontal
dashed line, whereas that of the RG theory ($1.586$)
-- by dot--dot--dashed line.}
\label{zeroref}
\end{figure}

The data points of $y_{eff}'$ look (and are expected to be) less accurate
than those of $y_{eff}''$, since $Re \left( u_1^0-u_c \right)$
has a very small value.
The $y_{eff}''$ data do not look scattered, therefore they allow a refined
analysis with account for nonlinear corrections. To obtain stable results,
we have included the data for smaller lattice sizes $L=3$ and $L=4$ given
in~\cite{ABV}. In principle, we can use rather arbitrary analytical function
$\phi (\beta)$ to evaluate the effective critical exponent
$$y_{eff}''(L) = \ln \left[ Im \, \phi \left(\beta_1^0(L/2) \right)
/ Im \, \phi \left( \beta_1^0(L) \right) \right] / \ln 2$$
and estimate its asymptotic value $y''$. For an optimal choice, however,
$y_{eff}''(L)$ vs $L^{-\omega}$ plot should be as far as possible linear
to minimize the extrapolation error. In this aspect, our choice
$\phi = \exp(-\beta)$ is preferable to $\phi = \exp(-4 \beta)$ used
in~\cite{ABV}. We have tested also another possibility, i.~e.
$\phi = \sinh( 2 \beta)$, which appears as a natural parameter in
the case of 2D Ising model.
The shape of the $y_{eff}''(L)$ plot can be satisfactory
well approximated by a third--order, but not by a second--order,
polinomial in $L^{-1/2}$, as it can be well seen when
analyzing the local slope of this curve. The
corresponding four parameter least--squares fits are shown in
Fig.~\ref{zeroref}.
They yield $y'' \simeq 1.473$ in the case of
$\phi = \exp(-\beta)$ (long--dashed line) and
$y'' \simeq 1.518$ at $\phi = \sinh(2 \beta)$ (solid line).
It is evident from Fig.~\ref{zeroref} that
in the latter case we have slightly better linearity of the fit,
therefore $1/\nu \simeq 1.518$ is our best estimate of 
the critical exponent $1/\nu$ from the actual MC data.
Thus, while the row estimation provided the value
$y''=1/\nu \simeq 1.552$ which is closer to the RG 
prediction $1/\nu \simeq 1.586$ (horizontal dot--dot--dashed line),
the refined analysis reveals remarkably better agreement with our
(exact) value $1/\nu=1.5$ (horizontal dashed line).

\section{$\lambda \varphi^4$ model and its crossover to Ising model}

Here we discuss a $\varphi^4$ model
on a three--dimensional cubic lattice. The Hamiltonian of this model,
further called $\lambda \varphi^4$ model, is given by
\begin{equation} \label{eq:H}
H/T= \sum\limits_{\bf x} \left\{ -2 \kappa \sum\limits_{\mu}
\varphi_{\bf x} \varphi_{{\bf x}+\hat \mu} + \varphi_{\bf x}^2
+ \lambda \left( \varphi_{\bf x}^2 -1 \right)^2 \right\} \;,
\end{equation}
where the summation runs over all lattice sites,
$T$ is the temperature, \linebreak $\varphi_{\bf x} \in \, ]-\infty; +\infty[$
is the scalar order parameter at the site with coordinate ${\bf x}$,
$\hat \mu$ is a unit vector in the $\mu$--th direction,
$\kappa$ and $\lambda$ are coupling constants.
Obviously, the standard 3D Ising model is recovered in the limit
$\lambda \to \infty$ where $\varphi_{\bf x}^2$ fluctuations
are suppressed so that, for a relevant configuration,
$\varphi_{\bf x}^2 \simeq 1$ or $\varphi_{\bf x} \simeq \pm 1$ holds.
The MC data for the Binder cumulant in this
$\lambda \varphi^4$ model have been interpreted in accordance with
the $\epsilon$--expansion and a perfect agreement
with the conventional RG values of critical exponents has
been reported in~\cite{Hasenbusch}.
According to the definition in~\cite{Hasenbusch}, the Binder cumulant
$U$ is given by
\begin{equation} \label{eq:U}
U= \frac{\langle m^4 \rangle}{ \langle m^2 \rangle^2} \;,
\end{equation}
where
$m=L^{-3} \sum_{\bf x} \varphi_{\bf x}$ is the magnetization
and $L$ is the linear size of the system.
Based on the $\epsilon$--expansion, it has been
suggested in~\cite{Hasenbusch} that, in the thermodynamic limit
$L \to \infty$, the value of the Binder cumulant
at the critical point $\kappa=\kappa_c(\lambda)$ and, equally, at
a fixed ratio $Z_a/Z_p=0.5425$ (the precise value is not important)
of partition functions with periodic
and antiperiodic boundary conditions is a universal constant $U^*$
independent on $\lambda$. We suppose that the latter statement
is true, but not due to the $\epsilon$--expansion.
It is a consequence of some general argument of the RG theory:
on the one hand, $U$ is invariant under the RG transformation and,
on the other hand, an unique fixed point (not
necessarily the Wilson--Fisher fixed point) exists in the case of
an infinite system, so that $U \equiv U^*$ holds at $L \to \infty$
and $\kappa = \kappa_c(\lambda)$ where $U^*$ is the fixed--point
value of $U$. The above conclusion remains true if we allow
that the fixed point is defined not uniquely in the sense that it
contains some irrelevant degree(s) of freedom
(like $c^*$ and $\Lambda$ in the perturbative RG theory discussed
in Sec.~2 of~\cite{K1}) not changing $U$.
The numerical results in~\cite{HV} confirm the idea that
$\lim_{L \to \infty} U(L) = U^*$ holds at criticality, where $U^*$
is a universal constant independent on the specific microscopic
structure of the Hamiltonian.

\section{Estimation of the correction exponent $\omega$}
\label{sec:omega}

\begin{figure}
\centerline{\psfig{figure=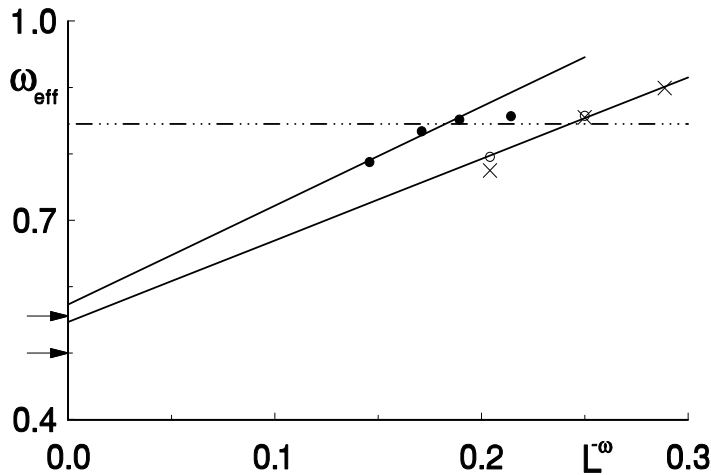,width=11cm,height=8.5cm}}
\vspace{-9.5ex}
\caption{\small Effective correction--to--scaling exponent
$\omega_{eff}(L)$ in the $O(n)$--symmetric $\lambda \varphi^4$ model
with $n=1$ (empty circles and crosses)
and $O(2)$--symmetric $dd-XY$ model (solid circles) depending on the
system size $L$. The linear least--squares fits give row estimates
of the asymptotic $\omega$ values $0.547$ and $0.573$,
at $n=1$ and $2$, respectively.
The corresponding theoretical values of the GFD theory
$1/2$ and $5/9$
(used in the $L^{-\omega}$ scale of the horizontal axis)
are indicated by arrows.
The dot--dot--dashed line shows the value $0.845(10)$ proposed
in~\cite{Hasenbusch} for the $3D$ Ising universality class ($n=1$).}
\label{om}
\end{figure}

Based on the idea that $U^*$ is constant for a given universality
class, here we estimate the correction--to--scaling exponent $\omega$.
According to Sec.~\ref{sec:crex}, corrections to finite--size scaling
for the magnetization of the actual 3D Ising and
$\lambda \varphi^4$ models are represented by an expansion in terms
of $L^{-\omega}$ where $\omega=1/2$. One expects that the
magnetization (Binder) cumulant~(\ref{eq:U}) has the same singular
structure. Since $\lim_{L \to \infty} U(L,\lambda) \equiv U^*$ holds at
a fixed ratio $Z_a/Z_p$, a suitable ansatz for estimation
of $\omega$ is~\cite{Hasenbusch}
\begin{equation} \label{omef}
U(L,\lambda_1)-U(L,\lambda_2) \simeq const \cdot L^{-\omega}
\hspace{4ex} \mbox{at} \hspace{2ex} Z_a/Z_p=0.5425 \;,
\end{equation}
which is valid for any two different nonzero values $\lambda_1$
and $\lambda_2$ of the coupling constant $\lambda$.
The data for $\Delta U(L)= U(L,0.8)-U(L,1.5)$ can be read
from Fig.~1 in~\cite{Hasenbusch} (after a proper magnification)
without an essential loss of the numerical accuracy, i.~e., within
the shown error bars. Doing so, we have evaluated the effective exponent
\begin{equation}
\omega_{eff}(L) = \ln \left[ \Delta U(L/2)/ \Delta U(L) \right] / \ln 2 \;,
\end{equation}
i.~e., $\omega_{eff}(12) \simeq 0.899$,
$\omega_{eff}(16) \simeq 0.855$, and $\omega_{eff}(24) \simeq 0.775$.
These values are shown in Fig.~\ref{om} by crosses.
Such an estimation,
however, can be remarkably influenced by the random scattering
of the simulated data points, particularly, at larger sizes where
$\Delta U(L)$ becomes small. This effect can be diminished if the
values of $\Delta U(L)$ are read from a suitable smoothened curve.
We have found that $\Delta U(L)$ within $L \in [7;24]$
can be well approximated by a second--order polinomial in $L^{-1/2}$.
The corresponding refined values $\omega_{eff}(16) \simeq 0.8573$ and
$\omega_{eff}(24) \simeq 0.7956$ read from this curve
are depicted in Fig.~\ref{om} by empty circles.
These values are similar to those obtained by a direct calculation
from the original data points (crosses).

In such a way, we see from Fig.~\ref{om}
that the effective exponent $\omega_{eff}(L)$ decreases
remarkably with increasing of $L$. According to GFD theory,
$\omega_{eff}(L)$ is a linear function of $L^{-1/2}$ at $L \to \infty$,
as consistent with the expansion in terms of $L^{-\omega}$ where
$\omega=0.5$. More data points,
including larger sizes $L$, are necessary for a reliable
estimation of the asymptotic exponent
$\omega =\lim_{L \to \infty} \omega_{eff}(L)$. Nevertheless, already
a row linear extrapolation in the scale of $L^{-1/2}$ 
with the existing data points yields the result $\omega \approx 0.547$
which is reasonably close to the exact value $0.5$ (horizontal dashed
line in Fig.~\ref{om}) found within
the GFD theory. The corresponding least--squares fit with circles
(at $L=24,16$) and cross (at $L=12$) is shown in Fig.~\ref{om} by
a straight solid line. It is evident from Fig.~\ref{om} that the final
result $\omega=0.845(10)$ (horizontal dot--dot--dashed line) reported
in~\cite{Hasenbusch} represents some average
effective exponent for the interval $L \in [6;24]$.
It has been claimed in~\cite{Hasenbusch} that the estimates for
$\omega$ (cf. Tab.~2 in~\cite{Hasenbusch}) are rather stable
with respect to a variation of $L_{min}$, where $L_{min}$ is the 
minimal lattice size used in the fit. Unfortunately, the analysis
has been made in an obscure fashion, i.~e., giving no original
data, so that we cannot check the correctness of this claim.
Besides, the estimates in Tab.~2 of~\cite{Hasenbusch} has been made
by using an ansatz
\begin{equation} \label{omef1}
U(L,\lambda) = U^* + c_1(\lambda) L^{-\omega}
\hspace{4ex} \mbox{at} \hspace{2ex} Z_a/Z_p=0.5425 \;,
\end{equation}
which is worse than~(\ref{omef}). Namely, (\ref{omef}) and (\ref{omef1})
are approximations of the same order, but~(\ref{omef1}) 
contains an additional parameter $U^*$ which is not known precisely. 
The results of an analysis with the ansatz~(\ref{omef}), reflected
in Tab.~5 of~\cite{Hasenbusch}, are not convincing, since
only very small values of $L_{min}$ (up to $L_{min}=6$) have been
considered.

In any case, we prefer to rely on that information we can
check, and it shows that the claim
in~\cite{Hasenbusch} that $\omega=0.845(10)$ holds with
$\pm 0.01$ accuracy cannot be correct,
since $\omega_{eff}(L)$ is varied in the first decimal place.

 We have made a similar estimation of $\omega$ for the dynamically
diluted $O(2)$--symmetric ($n=2$) $XY$ ($dd-XY$) model simulated
in~\cite{CHPRV} ($n=2$). In the case of
the $dd-XY$ model, parameter $D$ (cf.~Eq.(6) in~\cite{CHPRV})
plays the role of $\lambda$ in~(\ref{omef}). The data for the Binder
cumulant in Fig.~1 of~\cite{CHPRV} look rather accurate, i.~e., not
scattered. This enables us to estimate $\omega_{eff}$ just from
the data at $D=1.03$ and $D=\infty$ ($XY$ model).
The resulting values of $\omega_{eff}$
are depicted in Fig.~\ref{om} by solid circles. The scale of
$L^{-\omega}$ is used, where $\omega=5/9$ is our
theoretical value of the correction--to--scaling exponent at $n=2$
consistent with the general hypothesis proposed in Sec.~\ref{sec:crex}.
As we see, the solid circles can be well located on a smooth line
which, however, is remarkably curved at smaller sizes. Due to the
latter reason, we have used only the last three points (the largest
sizes) for the linear fit (solid line) resulting in an 
estimate $\omega \approx 0.573$ which comes close to our theoretical
value $\omega=5/9=0.555 \ldots$

In summary, the extrapolated $\omega$ values (Fig.~\ref{om})
in both cases $n=1$ and $2$ are reasonably close to our theoretical
values $1/2$ and $5/9$ indicated by arrows. Only a small
systematic deviation is observed. This, likely, is due to the error of
linear extrapolation: the $\omega_{eff}(L)$ plots have a
tendency to curve down slightly.
The conventional (RG) estimate $\omega \approx 0.8$ more or less
corresponds to effective exponents for currently simulated finite
system sizes, but not to the asymptotic exponents.

\section{Fitting the susceptibility data at criticality}
\label{sec:fit}

In this section we discuss some fits of MC data at criticality.
According to the finite--size scaling theory, the susceptibility
$\chi$ near the critical point is represented by an expansion
\begin{equation} \label{chi}
\chi= L^{2-\eta} \left( g_0(L/\xi)
+ \sum\limits_{l \ge 1} L^{-\omega_l} g_l(L/\xi) \right) \;,
\end{equation}
where $g_l(L/\xi)$ are the scaling functions, $\xi$ is the correlation
length of an infinite system, $\eta$ is the critical exponent
related to the $k^{-2+\eta}$ divergence of the correlation function
in the wave vector space at criticality, and $\omega_l$ are
correction--to--scaling exponents, $\omega_1 \equiv \omega$
being the leading correction exponent. The correlation length
diverges like $\xi \propto t^{-\nu}$ at $t \to 0$, where
$t=1-\beta/\beta_c$ is the reduced temperature. Thus,
for large $L$, in close vicinity of the critical point
where $tL^{1/\nu} \ll 1$ holds Eq.~(\ref{chi}) can be
written as
\begin{equation} \label{chi1}
\chi= a \, L^{2-\eta} \left( 1 + \sum\limits_{l \ge 1} b_l
L^{-\omega_l} + \delta(t,L) \right) \;,
\end{equation}
where $a=g_0(0)$ and $b_l=g_l(0)/g_0(0)$ are the amplitudes, and
$\delta(t,L)$ is a correction term which takes into account the
deviation from criticality. In the first approximation it reads
\begin{equation} \label{delta}
\delta(t,L) \simeq c \cdot tL^{1/\nu} \;,
\end{equation}
where $c$ is a constant.

\begin{figure}
\centerline{\psfig{figure=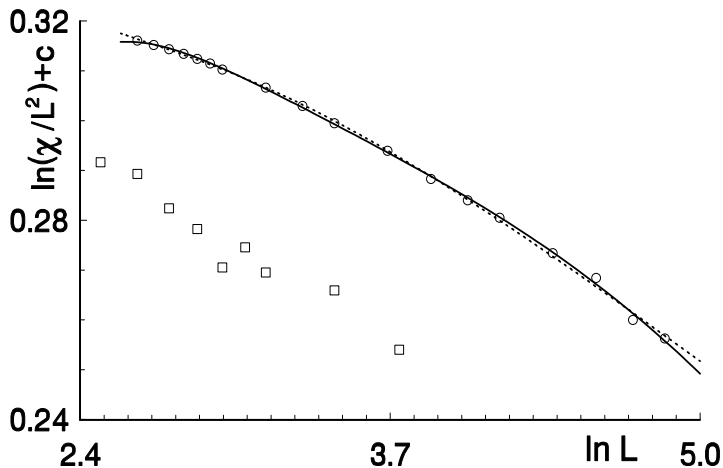,width=11cm,height=8.5cm}}
\vspace{-9.5ex}
\caption{\small The fits of $\ln \left( \chi/L^2 \right)$ data
at criticality (ansatz~(\ref{chi1})) shifted by a constant $c$.
Solid circles represent the MC data for 3D Ising model~\cite{HV} at
$\beta=0.2216545$ ($c=0$).
The fits with our (GFD) exponents 
($\ln a=1.065289, b_1=-2.72056, b_2=8.18636, b_3=-10.49614$)
and with those of~\cite{Hasenbusch,Justin1}
($\ln a=0.430933, b_1=0.05850, b_2=-7.74767, b_3=12.42890$)
are shown by solid and tiny--dashed lines, respectively.
The empty boxes are MC data for 3--component 3D $XY$ model~\cite{NhM},
shifted by $c=0.85$.} 
\label{chifi}
\end{figure}

 We start our analysis with the standard 3D Ising model with the Hamiltonian
\begin{equation}
H/T= -\beta \sum\limits_{\langle i j \rangle} \sigma_i \sigma_j \;.
\end{equation}
The critical point of this model has been found
in~\cite{HV} to be  $\beta_c \simeq 0.2216545$.
We have made our own tests with the data of~\cite{HV}, and have obtained
the same value within the uncertainty of $\pm 10^{-7}$.
From the maximal values of the derivative
$\partial \ln \langle m^2 \rangle / \partial \beta \equiv
\partial \ln \chi / \partial \beta$ evaluated in~\cite{FL} we conclude
that the shift of $\beta$ by $10^{-7}$ produces
the variation of $\ln \chi$ at $L=96$ near
$\beta=\beta_c$, which does not exceed $4.7 \cdot 10^{-4}$ in magnitude.
The latter means that, with a good enough accuracy, we may assume
that $\beta_c$ is just $0.2216545$ when fitting the susceptibility
data at criticality within $L \in [4;128]$. Here we mean the MC data
given in Tab.~25 of~\cite{HV}.
We have made and compared several fits of these data to 
ansatz~(\ref{chi1}) with $\delta(t,L)=0$
(more precisely, to the corresponding formula
for $\ln \chi$) for two different sets of the critical
exponents, i.~e., our (GFD) and that proposed in~\cite{Hasenbusch}.
The fits made with our exponents systematically
improve relative to those made with the exponents of~\cite{Hasenbusch},
as the system sizes grow and the approximation order increases.
The necessity to include several correction terms is dictated
by the fact that corrections to scaling are rather strong.
According to the least--squares criterion, the fit with our exponents
$\eta=1/8$ and $\omega_l=l/2$ becomes better than that provided
by the more conventional exponents $\eta=0.0358(4)$, $\omega_1=0.845(1)$,
$\omega_2=2 \omega_1$, and $\omega_3=2$~\cite{Hasenbusch} starting with
$L_{min}=28$ (i.~e., $L \in [L_{min};128]$), if two correction terms
($l=1,2$) are included.
In the case of three correction terms it occurs already at $L_{min}=11$.
The four--parameter ($a$, $b_1$, $b_2$, $b_3$) fits to MC data (empty
circles) within $L \in [14;128]$ are shown in Fig.~\ref{chifi}.
The fit with our exponents (upper solid line) is relatively better at
larger sizes. However, both fits (upper solid and dashed lines) look,
in fact, quite similar, so that we cannot make unambiguous conclusions
herefrom.

\begin{figure}
\centerline{\psfig{figure=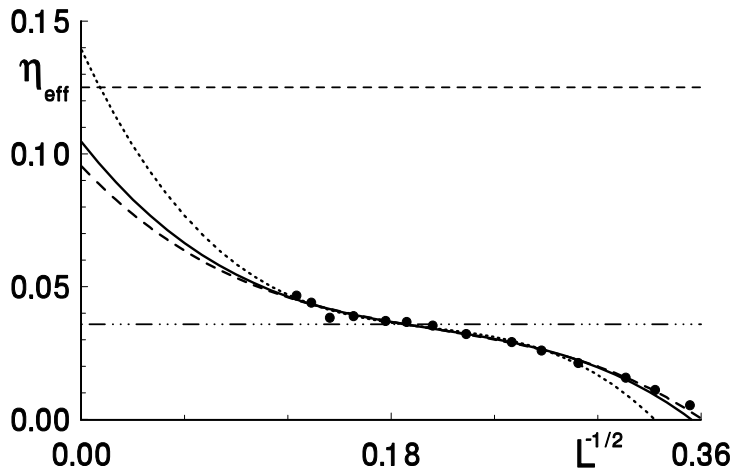,width=11cm,height=8.5cm}}
\vspace{-9.5ex}
\caption{\small The effective critical exponent $\eta_{eff}(L)$ (solid
circles) obtained by fitting the susceptibility data of 3D Ising model
at criticality ($\beta=0.2216545$)~\cite{HV} within the interval $[L;2L]$.
The least--squares approximations obtained by fitting the $\eta_{eff}(L)$
data within $[L_{min};64]$ to a third--order polinomial in $L^{-1/2}$ are
shown by dashed ($L_{min}=9$), solid ($L_{min}=10$), and tiny--dashed
($L_{min}=12$) lines. The asymptotic value $\eta=1/8$ of the GFD
theory is indicated by a horizontal dashed line. The dot--dot--dashed
line represents the $\eta$ value $0.0358$ proposed in~\cite{Hasenbusch}.}
\label{eteff}
\end{figure}

For comparison, we have shown in Fig.~\ref{chifi} also the MC data
for 3D $XY$ model~\cite{NhM}, where the order parameter is 3--component
vector with only two interacting components.
As we see, the actual MC data (empty boxes) at
$\beta_c$ evaluated approximately $\beta_c \simeq 0.6444$~\cite{NhM} are
rather scattered and, therefore, unsuitable for a refined analysis.
Nevertheless, this is a typical situation where authors of such data make
a very "accurate" and "convincing" estimation $\gamma/\nu=1.9696(37)$
or $\eta=0.0304(37)$ making a simple linear fit. However, the refined
analysis given above has shown that even in the case of 3D Ising model,
where the data are incompatibly more accurate, it is not so easy to
distinguish between $\eta=0.0358$ and $\eta=1/8$. Moreover, a refined
analysis prefer the second value which is much larger than those
usually provided by linear fits at typical system syzes $L \le 48$.
This is particularly well seen in Fig.~\ref{eteff}, where
the effective critical exponent $\eta_{eff}(L)$ of the
3D Ising model, estimated via the linear fit within $[L;2L]$, is depicted
by solid circles.
As we see, $\eta_{eff}(L)$ tends to increase well above
the conventional value $0.0358$ (horizontal dot--dot--dashed line).
The shape of the $\eta_{eff}(L)$ plot is satisfactory well reproduced
by a third--order polinomial in the actual scale of $L^{-1/2}$.
Three such kind of least--squares approximations
(at $L_{min}=9,10,12$) are shown in Fig.~\ref{eteff}.
These fits do not provide very accurate
and stable asymptotic values of $\eta$. Nevertheless, they are
more or less in agreement with our theoretical prediction
$\eta=1/8$ (horizontal dashed line). Besides, the values of $\eta_{eff}$
are affected by the error in $\beta_c$ (about $10^{-7}$) only slightly,
i.~e., by an amount not exceeding $0.001$.

\section{A test for 3D Ising model with "improved" action}
\label{imp}

Here we discuss some estimations of the critical exponents
from the susceptibility data of 3D Ising model, reported in~\cite{HV},
with the so called "improved" action (i.~e., $H/T$).
One of the problems with the standard 3D Ising model is that corrections
to scaling are strong. It has been proposed in~\cite{HV} to solve
this problem by considering a modified (spin--1) Ising model
with the Hamiltonian
\begin{equation} \label{Hsp1}
H/T= - \beta \sum\limits_{\langle i j \rangle}
\sigma_i \sigma_j + D \sum\limits_{i} \sigma_i^2  \;,
\end{equation}
where the spin $\sigma_i$ takes the values $0, \pm 1$,
with two coupling constants $\beta$ and $D$ adjusted in such a way that
the leading correction to
finite--size scaling vanishes for all relevant physical quantities
(magnetization cumulant, energy per site, susceptibility, etc.) and their
derivatives. Moreover, according to the claims in~\cite{HV}
(see the conclusions in~\cite{HV}), the ratios of
the leading and subleading corrections are universal, so that not only
the leading but all (!) corrections should vanish simultaneously.

\begin{figure}
\centerline{\psfig{figure=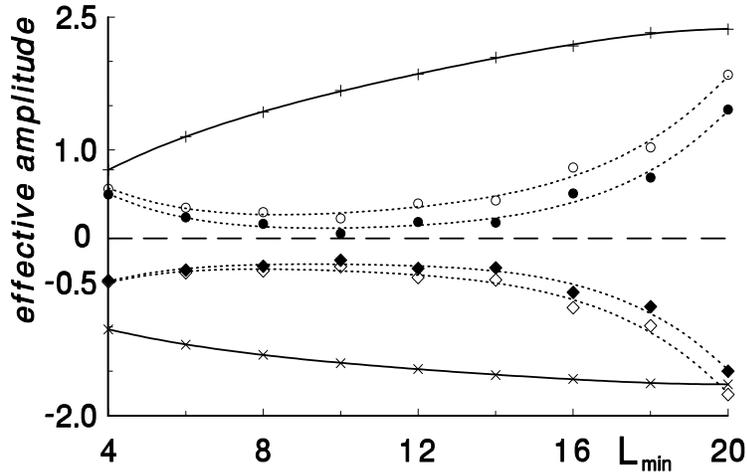,width=11cm,height=8.5cm}}
\vspace{-9.5ex}
\caption{\small The effective amplitudes $10 b_1$ (circles) and
$b_2$ (rhombs) in~(\ref{chi1}) estimated at fixed exponents
$\eta=0.0358$, $\omega_1=0.845$, $\omega_2=2 \omega_1$, and
$\nu=0.6296$ by fitting the MC data within $L \in [L_{min};56]$.
Filled symbols correspond to $\delta(t,L)=0$, empty
symbols -- to $\delta(t,L)=10^{-6} L^{1/\nu}$. The effective
amplitudes $b_1$ and $b_2$ estimated with the critical
exponents of our GFD theory ($\eta=1/8$, $\omega_l=l/2$) at
$\delta(t,L)=0$ are shown by "x" and "+", respectively.
Lines represent the least--squares approximations by a fourth--order
polinomial in $L$.}
\label{b}
\end{figure}

We have checked the correctness of these claims as described below.
We have fitted the corresponding to~(\ref{chi1}) expression for
$\ln \chi$ to the susceptibility data of the "improved" 3D Ising
model~(\ref{Hsp1}) with $(\beta,D)=(0.383245, 0.624235)$
(this is an approximation of the critical point) given
in~\cite{HV} (Tab.~26). By fixing the exponents, the least--squares
fit within $L \in [L_{min};56]$ (here $L=56$ is the maximal size
available in Tab.~26 of~\cite{HV}), including the leading and the
subleading correction to scaling, provides the effective amplitudes
$a$, $b_1$, and $b_2$ depending on $L_{min}$.
We have made a test with the critical exponents $\eta=0.0358(4)$,
$\omega=0.845(10)$, and $\nu=0.6296(3)$ proposed in~\cite{Hasenbusch}.
These values are close to those of the usual RG expansions~\cite{Justin1},
but, as claimed in~\cite{Hasenbusch}, they are more accurate.
According to~\cite{Hasenbusch}, the asymptotic expansion contains corrections
like $L^{-n \omega}$ and $L^{-2n}$, where $n=1, 2, 3, \ldots$
Thus we have $\omega_1=\omega$ and $\omega_2=2 \omega$.
The resulting amplitudes $10 b_1(L_{min})$ and $b_2(L_{min})$
are shown in Fig.~\ref{b} by circles and rhombs, respectively.
We have depicted by filled symbols
the results of the fitting with $\delta(t,L)=0$, assuming that the
critical coupling $\beta_c=0.383245$ has been estimated in~\cite{Hasenbusch}
with a high enough (6 digit) accuracy. The data points quite
well fit smooth (tiny dashed) lines within $L_{min} \in [4;20]$, which
means that the statistical errors are reasonably small.
If the exponents used in the fit are correct and corrections to
scaling are small indeed, then the convergence of the effective amplitudes
to some small values is expected with increasing of $L_{min}$.
However, as we see from Fig.~\ref{b}, the effective amplitudes tend
to increase in magnitude acceleratedly as $L_{min}$ exceeds $14$.
A small inaccuracy in $\beta_c$ value can be compensated by
the term $\delta(t,L) \simeq c^* L^{1/\nu}$ in~(\ref{chi1}),
where $c^*=ct$ (cf.~Eq.~(\ref{delta})). The results of fitting with
$c^*=10^{-6}$ are shown in Fig.~\ref{b} by empty symbols. As we see,
the expected inaccuracy in $\beta_c$ of order $10^{-6}$ does not
change the qualitative picture. The increase of the effective
amplitudes indicates that either the exponents are false, or
the asymptotic amplitudes are not small (or both). This is our
argument that the claims in~\cite{HV} about very
accurate critical exponents, extracted from the 3D Ising model
with "improved" action, are incorrect.

For comparison, we have shown in Fig.~\ref{b} also the
effective amplitudes $b_1(L_{min})$ and $b_2(L_{min})$
(by "x" and "+", respectively) estimated with the critical exponents of
our GFD theory (Sec.~\ref{sec:crex}) ($\eta=1/8$, $\omega_l=l/2$), assuming
$\delta(t,L)=0$.
The effective amplitudes converge to some values with increasing of
$L_{min}$. These, however, are not the true asymptotic values, since
the maximal size of the system has been eliminated to $L=56$.

\section{A test for the standard 3D Ising model}
\label{stan}

A test with the effective amplitudes, as in Sec.~\ref{imp},
appears to be more sensitive tool as compared to the fits discussed
in Sec.~\ref{sec:fit}. 
Here we consider the standard 3D Ising model. 
We have fitted all data points in Tab.~25 of~\cite{HV}
within the interval of sizes $[L;8L]$ 
to the theoretical expression
for $\ln \chi$ (consistent with~(\ref{chi1}))
to evaluate the effective amplitudes $a$ and $b_l$ with $l=1, 2, 3$
depending on $L$. Exceptionally in the case if all the
involved exponents are correct (exact) each effective amplitude can converge
to a certain nonzero asymptotic value at $L \to \infty$. In other words,
if one tries to compensate the inconsistency in the exponent by
choosing appropriate amplitude, then the amplitude
tends either to zero or infinity at $L \to \infty$.

\begin{figure}
\centerline{\psfig{figure=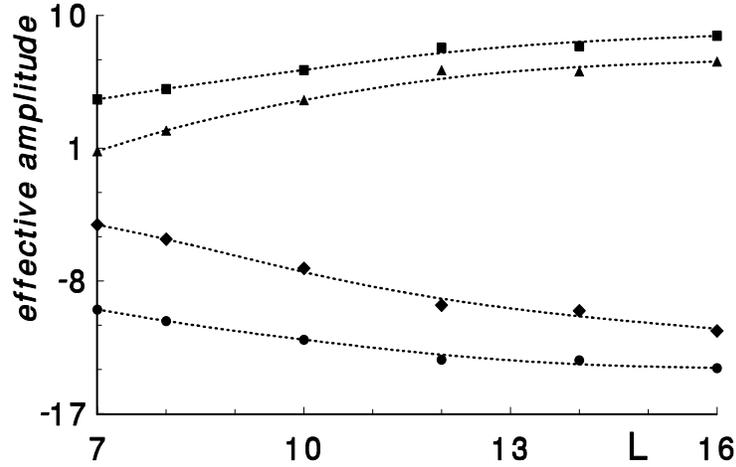,width=11cm,height=8.5cm}}
\vspace{-9.5ex}
\caption{\small The effective amplitudes in Eq.~(\ref{chi1})
$100 \, (\ln a(L) -1)$ (triangles), $5 b_1(L)$ (circles), $b_2(L)$ (squares),
and $b_3(L)$ (rhombs)
evaluated by fitting the susceptibility data of 3D Ising model at
criticality within the interval of sizes $[L;8L]$ with the critical
exponents $\eta=1/8$ and $\omega_l=l/2$ of the GFD theory.
}
\label{m}
\end{figure}

\begin{figure}
\centerline{\psfig{figure=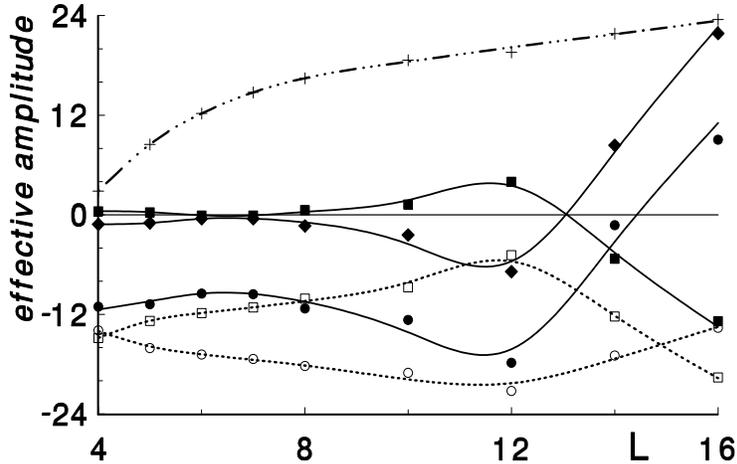,width=11cm,height=8.5cm}}
\vspace{-9.5ex}
\caption{\small The effective amplitudes in Eq.~(\ref{chi1})
evaluated by fitting the susceptibility data of 3D Ising model at
criticality within the interval of sizes $[L;8L]$ with the critical
exponents $\eta=0.0358$, $\omega_1=0.845$, $\omega_2=2 \omega_1$,
and $\omega_3=2$ proposed in~\cite{Hasenbusch}.
Solid symbols show the four--parameter fit:
$50 b_1(L)$ (circles), $b_2(L)$ (squares), and $b_3(L)$ (rhombs);
empty symbols show the three--parameter fit: $100 b_1(L)$ (circles) and
$27 b_2(L)$ (squares); crosses represent
the amplitude of the two--parameter fit, i.~e., quantity
$190 \, (b_l(L)+0.34)$.
}
\label{v}
\end{figure}

We have shown in Fig.~\ref{m} the effective amplitudes $\ln a(L)$ and
$b_l(L)$ in the case of our critical exponents $\eta=1/8$ and $\omega_l=l/2$.
As we expected, the effective amplitudes  
converge to some nonzero values with increasing of $L$.
This is a good numerical evidence that our critical exponets are true.
The case with the exponents of~\cite{Hasenbusch} $\eta=0.0358(4)$,
$\omega_1=0.845(10)$, $\omega_2=2 \omega_1$, and $\omega_3=2$ 
is illustrated in Fig.~\ref{v}.
As we expected, the effective amplitudes of our four--parameter fit
(solid symbols) tend to diverge with increasing of $L$,
which shows that this set of critical exponents is false.
One could object that, probably, the instability of the effective
amplitudes is due to small errors in MC data. However,
the amplitudes $b_1(L)$ and $b_2(L)$ of the more stable three--parameter
fit ($l=1,2$ in~(\ref{chi1})) behave in a similar way (see empty symbols in
Fig.~\ref{v}). Moreover, the amplitude $b_1(L)$ of the two--parameter fit,
shown by crosses, increases almost
linearly at large enough $L$ instead of the expected (in a case
of correct exponents) saturation like
$b_1(L) \simeq b_1 + const \cdot L^{-\omega}$.
As regards the convergence in Fig.~\ref{m} of the effective amplitudes
at $L \to \infty$, it is possible only if both conditions
are fulfilled, i.~e., the exponents are correct and the MC data
are accurate enough to ensure stable results. Thus, in any case,
the analysis in Fig.~\ref{m} provides rather convincing evidence
that our exponents are the true ones, which by itself rules
out the possibility that those proposed in~\cite{Hasenbusch}
could be correct.
The results in Figs.~\ref{m} and~\ref{v} are affected insignificantly
by a small inaccuracy of about $10^{-7}$ in the estimated $\beta_c$ value.

\section{Remarks about other numerical results}

There exists a large number of numerical results in the published
literature not discussed here and in~\cite{K1}.
A detailed review of these results is given in~\cite{PV}.
The cited there papers report results which disagree with
the values of the critical exponents we have proposed.
However, as regards the pure Monte Carlo study, we are quite confident
that, just like in the actually discussed case of 3D Ising model,
the increase of system syzes and/or use of higher--level
approximations will lead to the conclusion that fits with our
exponents are better than those with the conventional (RG) exponents.
Particularly, a careful analysis of the effective exponents made
in Secs.~\ref{sec:zeros}, \ref{sec:omega}, and~\ref{sec:fit}
already has shown that the effective exponents deviate from
the values predicted by the perturbative RG theory and converge
more or less to those of the GFD theory at $L \to \infty$.
Together with the analysis of the experiment with superfluid
$^4 He$~\cite{K1}, we have presented totally $5$ independent evidences of
such a behavior. Besides, an important prediction of exceptionally
our theory regarding the corrections to scaling is confirmed by the
exact-algorithm transfer matrix calculations in Sec.~\ref{sec:result}.

There exists some background for the conventional claims in the published
literature that all the usual methods give consistent results which appear
to be in a good agreement with the predictions of the perturbative
RG theory. The perturbation expansions of the RG theory, as well
as the techniques of high-- and low--temperature series expansion are
merely not rigorous extrapolation schemes which work not too close to
criticality.
As a result, these methods produce some pseudo or effective critical
exponents which, however, often provide a good approximation
just for the range of temperatures not too close to $T_c$ (critical
temperature) where these methods make sense and, therefore, agree with
each other.
According to the finite--size scaling theory, $t L^{1/\nu}$
is a relevant scaling argument, so that not too small values of the
reduced temperature $t$
are related to not too large sizes $L \sim t^{-\nu}$. Therefore, one
can understand that the MC results for finite systems
often can be well matched to the conventional critical exponents
proposed by high temperature (HT) and RG expansions.
If, however, the level
of MC analysis (i.~e., the level of approximations used) is increased,
then it turns out that the "conventional" critical exponents are
not valid anymore, as it has been demonstrated in the current paper.
It is because the "conventional" exponents are not
the asymptotic exponents. Correct values of the asymptotic exponents
have been found in~\cite{K1} considering suitable theoretical limits
instead of formal expansions in terms of $\ln k$ (at criticality, where
$k$ is the wave vector magnitude) or $\ln t$ (approaching criticality)
which are meaningless at $k \to 0$ and $t \to 0$. 
These formal expansions lie in the basis of the RG expansions for
the critical exponents.
One argue that $\ln k$ diverges weakly, therefore 
the expansions in powers of $\ln k$ can be treated.
This is a nonsense: any term like $k^{-\lambda}$ with $\lambda<0$
can be formally expanded in terms of $\ln k$, but therefore it
does not become less divergent. Moreover, not only the powers
of $k$ but almost any function can be expanded in terms
of $\ln k$, therefore it is impossible to dechiper what 
is hiden behind such formal expansions in reality.
This is a serious problem, since even an exponentially
small correction (at $k \to 0$) can give a nonvanishing contribution
to such a formal expansion (see examples in Sec.~2 of~\cite{K1}).
The problem is not only formal:
it has been proven in~\cite{K1} that
the assumption that the $\epsilon$--expansion works and provides
correct results at $k \to 0$ leads to an obvious contradiction in
mathematics (cf.~Sec.~2 in~\cite{K1}). This fact alone cannot be
compensated even by an infinite number of numerical evidences
supporting the "conventional" critical exponents coming from the
RG expansions.

Our arguments, based on the current numerical analysis, are the
following. First, the calculations by exact algorithms in 
Sec.~\ref{sec:result} confirm our theoretical prediction, but not
that of the perturbative RG theory. Second, we have proposed here a 
very sensitive method (i.~e., a study of effective amplitudes) which 
allows to test the consistency of a given set of
critical exponents with the MC data including several (in our case up to $3$)
corrections to scaling. We have applied this method to one of the recent
and most accurate numerical data for the susceptibility in
3D Ising model, and have got a confirmation that our critical
exponents are true. It would be not correct to doubt our results based on
less sensitive methods and lower--level approximations.

We prefer to rely just on the data of pure MC simulations becose of the
following reasons. The so called Monte Carlo RG (MCRG) method is not
free of assumptions related to approximate renormalization.
We would like only to mention that the MCRG study in~\cite{GT} of 3D Ising
systems of the largest (to our knowledge) available in 
literature sizes, i.~e. up to $L=256$,
has not revealed an excellent agreement with the usual predictions of
the perturbative RG. In particular, an estimate $\omega \approx 0.7$
has been obtained~\cite{GT} which is smaller than the usual (perturbative)
RG value $\approx 0.8$, but still is larger than the exact value $0.5$
predicted by the GFD theory. 
The high--temperature series cannot give more
precise results than those extracted from the recent most accurate MC
data, including the actual data of~\cite{HV}, since these series
diverge approaching the critical point. One approximates the divergent
series by a ratio of two divergent series (Pade approximation),
but it is never proven that such a method converges to the exact result.
Besides, the comparison to our calculations in 2D Ising model
via exact algorithms (Sec.~\ref{sec:correc}) shows that the HT series analysis
leads to misleading conclusions regarding such fine effects
as corrections to scaling. These effects are relevant for 3D models.
It is interesting to compare the MC and HT estimates of the
critical point for the standard 3D Ising model, i.~e.,
$\beta_c \simeq 0.2216545$ (MC)~\cite{HV} and
$\beta_c = 0.221659 +0.000002/-0.000005$ (HT)~\cite{SA}.
It is clear that the MC value is more accurate:
if we look in~\cite{HV}, where the estimation procedure is
well illustrated, we can see that $\beta_c$ is definitely
smaller than $0.221659$, and the error seems to be much
smaller than the difference between both estimates $0.0000045$.
As we have mentioned already, our independent tests suggest
that the error of the actual MC value is about $10^{-7}$.

\section{Conclusions}

Summarizing the present work we conclude the following:
\begin{enumerate}
\item
Critical exponents and corrections to scaling for different physical
quantities have been discussed in framework of our~\cite{K1} recently 
developed GFD (grouping of Feynman diagrams) 
theory (Sec.~\ref{sec:crex}).

\item
Calculation of the two--point correlation function
of 2D Ising model at the critical point has been made
numerically by exact transfer matrix algorithms 
(Secs.~\ref{sec:algorithm} and~\ref{sec:correc}).
The results for finite lattices including up to
800 spins have shown the existence of a nontrivial
correction to scaling with a very small amplitude and
exponent about $1/4$ in agreement with the prediction
of our GFD theory. No correction with the conventionally
predicted exponent $4/3$ has been detected.

\item 
The recently published Monte Carlo data for several 
three--dimensional lattice models have been reanalyzed. 
This analysis in Secs.~\ref{sec:zeros} to~\ref{sec:fit}
has shown that the effective critical exponents deviate from
the values predicted by the perturbative RG theory and converge
towards those of the GFD theory at $L \to \infty$.
The same behavior has been observed in the experiment with
superfluid $^4 He$ discussed in~\cite{K1}. Totally, these are
five independent evidences of such a behavior, suggesting that the
above examples are not occasional or exceptional,
but reflect a general rule.

\item
Different sets of critical exponents (one provided by GFD theory, another
proposed in~\cite{Hasenbusch}) predicted for the 3D Ising model
have been tested by analyzing the effective amplitudes
(Sec.~\ref{imp} and~\ref{stan}).
While the usual fits of the susceptibility data do not allow to show
convincingly which of the discussed here sets of the critical exponents
is better, this method strongly suggests that the conventional
critical exponents $\eta=0.0358(4)$ and $\omega=0.845(10)$~\cite{Hasenbusch}
are false, whereas our (GFD) values $\eta=1/8$ and $\omega=1/2$ are true. 
\end{enumerate}

\section*{Acknowledgements}

This work including numerical calculations of the 2D Ising model 
have been performed during my stay at the Graduiertenkolleg 
\textit{Stark korrelierte
Vielteilchensysteme} of the Physics Department, Rostock University,
Germany.

\end{document}